\documentclass[12pt]{article}
\pdfoutput=1
\usepackage{jheppub}
\usepackage{amsmath}
\usepackage{amsfonts}
\usepackage{amssymb}
\usepackage{graphicx}
\usepackage{bm}

\usepackage[export]{adjustbox}
\DeclareMathOperator{\U}{U}
\DeclareMathOperator{\SU}{SU}
\DeclareMathOperator{\SO}{SO}

\newcommand{\de}{\partial}

\newcommand{\coma}{\text{ , }}
\newcommand{\fstop}{\text{ .}}

\newcommand{\e}{\text{ and }}
\newcommand{\AdS}{\text{AdS}}

\newcommand{\bmat}{\left(\begin{array}}
\newcommand{\emat}{\end{array}\right)}

\def\yzero{\smash{\hbox{$y\kern-4pt\raise1pt\hbox{${}^\circ$}$}}}

\def\beq{\begin{equation}}
\def\eeq{\end{equation}}
\def\beqa{\begin{eqnarray}}
\def\eeqa{\end{eqnarray}}

\def\-{\hphantom{-}}

\def\s2{\frac{1}{\sqrt2}}

\def\beq{\begin{equation}}
\def\eeq{\end{equation}}
\def\beqa{\begin{eqnarray}}
\def\eeqa{\end{eqnarray}}

\def\diag{{\rm diag \,}}
\def\IF{\relax{\rm I\kern-.18em F}}
\def\II{\relax{\rm I\kern-.18em I}}

\def\Dsl{\,\raise.15ex\hbox{/}\mkern-13.5mu D} 

\def\IC{{\bf{C}}}
\def\IS{{\bf {S}}}

\def\IZ{{\bf {Z}}}
\def\IX{{\bf {X}}}

\def\IT{{\bf {T}}}
\def\IP{{\bf {P}}}

\catcode`\@=11   
\newdimen\@rotdimen
\newbox\@rotbox  

\def\@vspec#1{\special{ps:#1}}
\def\@rotstart#1{\@vspec{gsave currentpoint currentpoint translate
   #1 neg exch neg exch translate}}
\def\@rotfinish{\@vspec{currentpoint grestore moveto}}
\def\@rotr#1{\@rotdimen=\ht#1\advance\@rotdimen by\dp#1%
   \hbox to\@rotdimen{\hskip\ht#1\vbox to\wd#1{\@rotstart{90 rotate}%
   \box#1\vss}\hss}\@rotfinish}
\def\@rotl#1{\@rotdimen=\ht#1\advance\@rotdimen by\dp#1%
   \hbox to\@rotdimen{\vbox to\wd#1{\vskip\wd#1\@rotstart{270 rotate}%
   \box#1\vss}\hss}\@rotfinish}%
\def\@rotu#1{\@rotdimen=\ht#1\advance\@rotdimen by\dp#1%
   \hbox to\wd#1{\hskip\wd#1\vbox to\@rotdimen{\vskip\@rotdimen
   \@rotstart{-1 dup scale}\box#1\vss}\hss}\@rotfinish}%
\def\@rotf#1{\hbox to\wd#1{\hskip\wd#1\@rotstart{-1 1 scale}%
   \box#1\hss}\@rotfinish}%
\def\rotate{\@ifnextchar[{\@rotate}{\@rotate[l]}}
\def\@rotate[#1]#2{\setbox\@rotbox=\hbox{#2}\@nameuse{@rot#1}\@rotbox}

\catcode`\@=12

\begin{document}

\makeatletter
\@addtoreset{equation}{section}
\makeatother
\renewcommand{\theequation}{\thesection.\arabic{equation}}
\pagestyle{empty}
\rightline{IFT-UAM/CSIC-20-46}
\vspace{1.2cm}
\begin{center}
\LARGE{\bf Discrete Symmetries, \\Weak Coupling Conjecture \\
and Scale Separation in AdS Vacua}
\\[12mm] 

\large{Ginevra Buratti, Jos\'e Calder\'on,  Alessandro Mininno, Angel M. Uranga\\[4mm]}
\footnotesize{Instituto de F\'{\i}sica Te\'orica IFT-UAM/CSIC,\\[-0.3em] 
C/ Nicol\'as Cabrera 13-15, 
Campus de Cantoblanco, 28049 Madrid, Spain}\\ 
\footnotesize{\href{mailto:ginevra.buratti@uam.es}{ginevra.buratti@uam.es}, \href{mailto:j.calderon.infante@csic.es}{j.calderon.infante@csic.es}, \href{mailto:alessandro.mininno@uam.es}{alessandro.mininno@uam.es},  \href{mailto:angel.uranga@csic.es}{angel.uranga@csic.es}}

\vspace*{10mm}

\small{\bf Abstract} \\
\end{center}
\begin{center}
\begin{minipage}[h]{\textwidth}
We argue that in theories of quantum gravity with discrete gauge symmetries, 
e.g. $\text{\textbf{Z}}_k$, the gauge couplings of U$(1)$ gauge symmetries 
become weak in the limit of large $k$, as $g\to k^{-\alpha}$ with $\alpha$ a 
positive order 1 coefficient. The conjecture is based on black hole arguments 
combined with the Weak Gravity Conjecture (or the BPS bound in the supersymmetric 
setup), and the species bound. We provide explicit examples based on type IIB 
on AdS$_5\times \text{\textbf{S}}^5/\text{\textbf{Z}}_k$ orbifolds, and M-theory on AdS$_4\times\text{\textbf{S}}^7/\text{\textbf{Z}}_k$ ABJM orbifolds 
(and their type IIA reductions). We study AdS$_4$ vacua of type IIA on CY 
orientifold compactifications, and show that the parametric scale separation in 
certain infinite families is controlled by a discrete $\IZ_k$ symmetry for 
domain walls. We accordingly propose  a refined version of the strong AdS 
Distance Conjecture, including a parametric dependence on the order of the 
discrete symmetry for 3-forms.
\end{minipage}
\end{center}
\newpage
\setcounter{page}{1}
\pagestyle{plain}
\renewcommand{\thefootnote}{\arabic{footnote}}
\setcounter{footnote}{0}

\tableofcontents

\vspace*{1cm}

\newpage

\section{Introduction and Conclusions}
By now there is a substantial amount of swampland conjectures constraining effective field theories to be compatible with Quantum Gravity \cite{Vafa:2005ui,Ooguri:2006in,ArkaniHamed:2006dz,Ooguri:2016pdq,Obied:2018sgi,Buratti:2018onj} (see \cite{Brennan:2017rbf,Palti:2019pca} for reviews). They have led to interesting insights into phenomenological applications of string theory models.

 Interestingly, many of these works focus on the properties of continuous gauge 
 symmetries, whereas far fewer results have been obtained to constrain discrete 
 symmetries (for some results, see 
 \cite{Banks:2010zn,Harlow:2018jwu,Harlow:2018tng}, and also \cite{Craig:2018yvw}), 
 and mostly focus on the constrain that global discrete symmetries, just like 
 global continuous symmetries, are forbidden in quantum gravity (see 
 \cite{Krauss:1988zc,Alford:1988sj,Alford:1989ch,Alford:1990mk,Alford:1991vr,Alford:1992yx,Alford:1990pt}
  for early literature). Discrete gauge symmetries are an interesting area with 
 exciting applications in BSM phenomenology and string model building 
 \cite{BerasaluceGonzalez:2011wy,BerasaluceGonzalez:2012zn,BerasaluceGonzalez:2012vb,Ibanez:2012wg,Marchesano:2014mla,Montero:2015ofa,Buratti:2018xjt}.
  The scarcity of swampland constraints on them is partially explained by the 
 fact that discrete symmetries lack long-range fields or tunable parameters 
 like coupling constants, so there are less handles to quantitatively constrain 
 their properties or their impact on other quantities of the theory. 

In this work, we overcome this difficulty by considering theories with both 
discrete and continuous gauge symmetries, and uncover interesting quantitative 
links among them. For simplicity we focus on abelian $\IZ_k$ and $\U(1)$ 
symmetries. In theories with a $\U(1)$ gauge symmetry, considerations about 
evaporation of charged black holes lead to the Weak Gravity Conjecture 
\cite{ArkaniHamed:2006dz}, by demanding that the black hole should remain 
(sub)extremal throughout the process. To put it simply, considering an extremal 
black hole with $M=gQ$ (in Planck units), the theory must contain particles 
with mass $m$ and charge $q$, with $m\leq gq$, such that the black hole can 
decay without becoming super-extremal. This is the Weak Gravity Conjecture 
(WGC). The marginal case in which the Weak Gravity Conjecture particles 
saturate the inequality $m=q$ has been further proposed to correspond to 
supersymmetric situations, in which it often corresponds to a BPS bound.

If the theory enjoys a further $\IZ_k$ discrete gauge symmetry, one can 
consider any such classical black hole solution and endow it with discrete 
$\IZ_k$ charge, with no change in the classical solution, as this charge does 
not source long-range fields  (see e.g. \cite{Coleman:1991ku}, and also \cite{GarciaGarcia:2018tua} for a recent perspective), and study their decay as in the WGC. In 
particular, we may consider extremal black holes carrying $\IZ_k$ charge and 
derive a striking result, the $\IZ_k$ Weak Coupling Conjecture (WCC) which 
schematically is the statement that in a theory with a discrete $\IZ_k$ gauge symmetry and a $\U(1)$ gauge symmetry with coupling $g$, the gauge coupling scales as $g\sim k^{-\alpha}$ for large $k$, with $\alpha$ a positive order 1 coefficient.

The derivation and some qualifications on this statement are discussed in Section \ref{sec:wcc}. In particular, we also relate this statement with diverse versions of swampland distance conjectures. 

As we will see, the derivation is most precise in the supersymmetric case, in 
which the WGC bound saturates, but we believe it holds far more generally, as 
we will illustrate in concrete string theory examples. In particular, in 
Section \ref{sec:ads5} we study $\AdS_5\times \IS^5/\IZ_k$ vacua (and 
generalizations to general toric\footnote{By toric, in this context we mean 
that the CY3 obtained as the real cone over the Sasaki-Einstein $5$d variety, is 
toric.} theories $\AdS_5\times \IX_5/\IZ_k$), in which there is a discrete 
Heinsenberg group ${\bf H}_k$, associated to torsion classes in $\IS^5/\IZ_k$ 
\cite{Gukov:1998kn,Burrington:2006uu,Garcia-Valdecasas:2019cqn}. This is 
generated by elements $A$, $B$, each generating a $\IZ_k$ symmetry, with 
commutation relations $AB=CBA$, with $C$ a central element. In the effective $5$d 
theory (namely at scales below the KK scale, and thus at long distance compared 
with the AdS radius as well) there is at least one $\U(1)$ gauge symmetry, 
corresponding to the R-symmetry of the holographic dual SCFT, whose coupling, as we 
show, obeys the WCC. In addition, for $\IS^5/\IZ_k$, and in fact for any 
toric theory $\IX_5/\IZ_k$, there are two additional $\U(1)$'s (the mesonic 
global symmetries in the dual SCFT), which also satisfy the WCC.

In Section \ref{sec:ads4} we discuss an analogous exercise in $4$d by considering in Section \ref{sec:mtheory} the case of M-theory on $\AdS_4\times \IS^7/\IZ_k$, which provides the gravity dual to the ABJM theories \cite{Aharony:2008ug}. The $\U(1)$ symmetry corresponds to an isometry of the internal space, and the discrete symmetry is also related to torsion classes in $\IS^7/\IZ_k$, although it has an intricate structure not reducible to just $\IZ_k$. This is further clarified using the type IIA perspective in Section \ref{sec:typeiia}, in which the discrete gauge symmetry is shown to have order $k^2+N^2$, and the $\U(1)$ symmetry is a linear combination of different RR p-form gauge symmetries, with a second linear combination that is massive due to a St\"uckelberg coupling. We discuss these systems and show how the corresponding WCC is duly satisfied.

In Section \ref{sec:iiacy} we turn to exploiting these considerations in 
theories in which the $\IZ_k$ charged objects are not particles (or their dual 
objects, e.g. strings in $4$d), but rather $4$d domain walls. In particular, we 
consider the type IIA $\AdS_4$ vacua obtained in CY orientifold compactification 
with NSNS and RR fluxes. In Section \ref{sec:dgkt} we review a class of 
compactifications with  fluxes scaling with a parameter $k$, shown in 
\cite{DeWolfe:2005uu} to have parametric scale separation controlled by $k$. 
These vacua would violate the strong AdS Distance Conjecture proposed in 
\cite{Lust:2019zwm}, an issue on which our analysis sheds important insights. In 
Section \ref{sec:dks} we show that these systems are higher p-form analogues 
to the type IIA vacua of Section \ref{sec:typeiia}, with a continuous 3-form 
symmetry arising from a massless linear combination, and the discrete symmetry 
arising from a second linear combination made massive by a 3-form St\"uckelberg 
mechanism (see \cite{Dvali:2005an,Kaloper:2008fb}, also \cite{Marchesano:2014mla}), also called Dvali-Kaloper-Sorbo 
(DKS) mechanism. In Section \ref{sec:scaling-from-symms} we discuss the role of 
the discrete $\IZ_k$ symmetry in fixing the scaling of the moduli with $k$. In 
Section \ref{sec:domain-walls} we use tensions of BPS domain walls to recover 
the vacuum energy scalings, and show that AdS vacua with trivial 3-form 
discrete symmetry have no scale separation, while the above scaling family 
of AdS vacua with a non-trivial 3-form discrete symmetry displays scale 
separation controlled by $k$, as follows. The scale separation relation between 
the KK scale $m_{KK}$ and the $4$d cosmological constant $\Lambda$ is given by 
the species bound
\beqa
\Lambda\, =\, \frac{m_{\rm KK}^{\, 2}}{k} \fstop
\eeqa
We accordingly formulate the following $\IZ_k$ Refined  Strong $\AdS_4$ Distance Conjecture: In supersymmetric $\AdS_4$ vacua with a discrete symmetry associated to $\IZ_k$-charged domain walls, the ratio between the KK scale and $\Lambda$ is $m_{\rm KK}\sim (k\Lambda)^{1/2}$.

This provides an underlying rationale for the seeming violation of the strong ADC by the family of scaling AdS solutions in type IIA vacua with field strength fluxes. It would be interesting to test it in other setups, and even exploit it in applications to holography.

\medskip

Our work is an important step in understanding the nature of discrete gauge symmetries in quantum gravity, and their non-trivial interplay with continuous gauge symmetries. As in other swampland constraints, although the arguments for the $\IZ_k$-WCC are admittedly heuristic, there is a substantial amount of evidence from concrete, very rigorous, string vacua supporting it. We have argued that discrete symmetries for 3-forms play an important role in the problem of scale separation, and provided a rationale to embed it in a refined AdS Distance Conjecture. We thus expect they may be relevant in other swampland criteria, like the de Sitter constraint. We hope to report on these topics in the near future.

\smallskip

\paragraph{Note:} As we were finishing writing this paper, ref. \cite{Junghans:2020acz} appeared, which studies scale separation in type IIA AdS vacua, albeit from a different perspective (note also \cite{Marchesano:2020rnd}, appeared shortly after our work). It would be interesting to explore the relation between the two approaches.

\section{The $\IZ_k$ Weak Coupling Conjecture}
\label{sec:wcc}

In this Section we consider theories of quantum gravity with discrete and continuous gauge symmetries. For simplicity we focus on a $\IZ_k$ discrete symmetry and a $\U(1)$ gauge symmetry. Generalizations to multiple $\U(1)$'s and discrete groups could be worked out similarly.
Notice that throughout the paper we are interested in the properties of the theory at large $k$, hence many of our expressions should be regarded as the leading approximation in a $1/k$ expansion.

\subsection{A black hole argument}
\label{sec:bh}

For concreteness we focus on $4$d theories, although the results extend to other dimensions (as we will see e.g. in the examples of Section \ref{sec:ads5}). The strategy is to use black hole evaporation as a guiding principle to derive new swampland constraints, as we now review in two familiar situations.

\subsubsection{Review of some mass bound derivations}
\label{sec:dvali-wgc}

Let us briefly recall one such derivation for the Weak Gravity Conjecture (WGC) 
\cite{ArkaniHamed:2006dz}. The idea is to consider extremal black holes, with 
mass $M$ and charge $Q$, satisfying $M=gQM_p$, where $g$ is the $\U(1)$ gauge 
coupling (in units in which the minimal charge is 1). Requiring the decay of 
such extremal black holes, while preventing them from becoming super-extremal, 
leads to the familiar statement of the Weak Gravity Conjecture, namely that 
there must exist some particle in the theory with mass $m$ and charge $q$ such 
that
\beqa
m\,\leq\, g\, q\, M_p.
\eeqa
There are different versions of the WGC (see \cite{Palti:2019pca} for a review with references), including the lattice \cite{Heidenreich:2015nta} and sublattice \cite{Heidenreich:2016aqi} versions, but we stick to the basic one above.

Let us consider a black hole (possibly charged under the $\U(1)$ or not),
carrying a discrete $\IZ_k$ charge. The analysis now follows
\cite{Dvali:2007hz}. Even though this is a gauge symmetry, it does not have
long-range fields, so it does not affect the classical black hole solution,
neither its evaporation in the semiclassical approximation, which thus does not
allow to eliminate the $\IZ_k$ charge. Since we are interested in the large $k$
behavior, this would lead to a too large number of remnants. Hence, when the
black hole radius reaches some cutoff value $\Lambda^{-1}$ it starts peeling off
its $\IZ_k$ charge. If we denote by $m$ the mass of the $\IZ_k$ charged
particles, the mass of the black hole at the cutoff scale should suffice to emit
${\cal O}(k)$ of such particles, that is
\beqa
M_{p}^{\, 2} \, \Lambda^{-1} \, \gtrsim\, k\, m\fstop
\eeqa
The cutoff radius is intuitively of the order of the inverse mass of the 
emitted particle, hence we consider $\Lambda\sim  \beta m$, with $\beta$ some 
unknown coefficient encoding model dependent information about the black hole 
and its evaporation process. Consequently, we obtain
\beqa
m^2\, \lesssim \, \frac{M_p^{\,2}}{k}\fstop
\label{dvali}
\eeqa
This is often known as the species bound \cite{Dvali:2007hz}, although in the present context $k$ does not correspond to the number of species, rather it relates to the order of the discrete symmetry.\footnote{Actually, to account for the fact that the particle needs not be minimally charged under $\IZ_k$, we should point out that the role of $k$ above should actually be played by the number of emitted particles. Hence the factor appearing in relations like (\ref{dvali}) may differ from the order of the discrete group by a factor of the particle charge, see some examples in Sections \ref{sec:ads4}, \ref{sec:iiacy}.}

Keeping in mind the unknown factors in the discussion, we take the above relation as controlling the scaling of suitable $\IZ_k$ charged particles in the limit of large $k$. Namely, there must exist some $\IZ_k$ charged particle whose mass must scale as $m\lesssim k^{-1/2} M_p$. 

In the following, we will apply this constraint to  black holes charged under 
continuous $\U(1)$ symmetries. One may worry that the derivation in 
\cite{Dvali:2007hz} did not include such charges, i.e. it implicitly assumed 
Schwarzschild black holes. However, there are analogous arguments for charged 
(in fact extremal) black holes in theories with $\U(1)$ gauge groups, leading 
to identical results, as we discuss in Appendix \ref{app:extremal}. Hence for 
practical purposes we may continue with the above simple picture.

\subsubsection{The $\IZ_k$ Weak Gravity Conjecture}
\label{Zk-WGC}

In the above discussion, the mass of the $\IZ_k$ particle we are constraining is thought of as the lightest one. However, in the following we argue that we can use a similar argument to constrain not only the lightest $\IZ_k$ charge particle, but also the one with smallest ratio $q/m$ between its $\U(1)$ charge and its mass. Namely, the Weak Gravity Conjecture particles.

Consider an extremal black hole with mass $M$ and charge $Q$, and endow it with a large $\IZ_k$ charge. The black hole can try to peel off its $\IZ_k$ charge by emitting $\IZ_k$ charged particles, but this would decrease its mass while keeping its charge fixed, thus becoming super-extremal. The simplest way to prevent this is that there exist some $\IZ_k$ charged particle which is also charged under the $\U(1)$ with charge $q$, and such that it satisfies the WGC bound $m\leq gqM_p$. In other words, the simplest resolution is that the WGC particles carry $\IZ_k$ charge. We may dub this result as the {\bf $\IZ_k$ Weak Gravity Conjecture}.

This is a remarkable result, but is actually a little bit of an overstatement. It may well happen that the WGC particles are neutral and do not saturate the WGC bound, and the evaporation of the black hole by emission of WGC particles makes it sufficiently sub-extremal so as to be able to subsequently emit enough $\IZ_k$ charged particles (not obeying the WGC bound) to peel off its discrete charge without ever getting super-extremal. Interestingly, notice that this is only possible if the WGC particles satisfy the strict WGC bound, not the equality, and hence, according to the extended WGC version in \cite{Ooguri:2016pdq}, it is possible only in non-supersymmetric theories. Thus our derivation above is strictly valid in the supersymmetric setup, and in our examples we will indeed focus on supersymmetric examples. We however still consider the argument as interestingly compelling also in non-supersymmetric models, and hence keep an open mind about its general validity, and that of its implications, to which we turn.

\subsubsection{The $\IZ_k$ Weak Coupling Conjecture}

The fact that the WGC particles, whose defining feature has to do with the $\U(1)$ gauge symmetry, know about the $\IZ_k$ symmetry implies that there are cross constraints among the $\U(1)$ and the $\IZ_k$ symmetry. Indeed, let us consider a relaxed version of the $\IZ_k$ bound (\ref{dvali}), by stating that the $\IZ_k$ charged particles involved in the black hole decay should have mass scaling as
\beqa
m\, \sim\, k^{-\alpha}\, M_p\coma
\eeqa
with $\alpha$ an order 1 coefficient, obeying some bound $\alpha \geq 1/2$ to 
satisfy (\ref{dvali}). On the other hand, the particles that extremal black 
holes use to peel off their $\IZ_k$ charge are WGC particles, hence obey
\beqa
m\,\sim\, g\, q\, M_p\fstop
\eeqa
We thus obtain that the gauge coupling of the $\U(1)$ must depend on $k$ and should become weak fast enough in the large $k$ limit, as
\beqa
g\sim k^{-\alpha}\fstop
\eeqa
We thus propose this to be a general swampland constraint, as follows:

\bigskip

\paragraph{$\IZ_k$ Weak Coupling Conjecture} {\em  In a quantum gravity theory with a discrete $\IZ_k$ gauge symmetry and a $\U(1)$ gauge symmetry with coupling $g$, the gauge coupling scales as $g\sim k^{-\alpha}$ for large $k$, with $\alpha$ a positive order 1 coefficient.}

\bigskip

This intertwining between the properties of discrete and continuous symmetries is completely unexpected from the viewpoint of the low energy effective field theory, where these parameters are uncorrelated and would seem to be completely free choices. As with other swampland constraints, it is amusing that quantum gravity manages to impose its own plans.

A simple illustration of how this interplay works in intersecting brane modes is discussed at the heuristic level in Appendix \ref{app:intersecting}. More concrete examples will follow in the upcoming sections.

\subsection{Distance Conjectures}
\label{sec:distance}

Before moving to concrete examples, it is interesting to explore the relation between the $\IZ_k$ WCC and the Swampland Distance Conjectures (SDC). The WCC states that gauge couplings scale to zero for large $k$, thus approaching a global symmetry and hence presumably leading to the appearance of a tower of states becoming light. 

An intuitive picture of this implication is as follows. Consider a $4$d version of the $\IZ_k$ WCC with $g\sim k^{-\alpha}$. For simplicity, and following many examples in string theory we consider $g$ to belong to a complex modulus 
\beqa
S=\frac{1}{g^2}+i\theta
\eeqa
and assume a K\"ahler potential 
\beqa
K(S, {\bar S})=-\log(S+{\bar S})\fstop
\eeqa
In this moduli space, the distance as a function of $s=\operatorname{Re}\,S$ as one approaches infinity reads
\beqa
d\sim \int \frac {ds}{s}\sim \log s\fstop
\eeqa
The SDC states that there is a tower of states becoming light as $s\to \infty$ with masses
\beqa
m_{\rm tw}\sim M_{p} \, e^{-\gamma d}\coma
\eeqa
with $\gamma$ an order 1 coefficient, for $d$ measured in Planck units. In our case we have
\beqa
m_{\rm tw} \sim \, M_p\, k^{-\frac 12\alpha \gamma}\fstop
\eeqa
Hence there is a $\IZ_k$ Distance Conjecture stating that there is a tower of states with masses becoming light as a negative power of $k$. This is just a re-derivation of the `species' bound cutoff \cite{Dvali:2007hz}.

\medskip

In fact, the above argument where $g$ is dealt with as a modulus going to infinite distance in moduli space does not correspond to the general $\IZ_k$ WCC, since at least some of the gauge couplings may not correspond to fundamental moduli. For instance, consider the intersecting brane toy model in Appendix \ref{app:intersecting}. There, the moduli remain at fixed location in moduli space, and we instead change the discrete wrapping numbers for some D-branes. Hence, the origin of the tower should be a different one, as is easily argued. In a configuration in which one stack of branes has wrappings scaling with $k$, the angles between that stack of branes and others will scale as $\theta\sim k^{-1}$ (to see that, consider e.g. the cycles $(1,0)$ and $(k,1)$ in a rectangular $\IT^2$ with radii $(R_1,R_2)$. They have intersection angle $\theta$ with $\tan\theta=k^{-1}R_2/R_1$, hence $\theta\sim k^{-1}$). As discussed in \cite{Aldazabal:2000cn,Aldazabal:2000dg} there is a tower of string states with masses given by 
\beqa
m_{\rm tw}^2\sim M_s\,\theta\sim  k^{-1}\fstop
\eeqa
This again nicely reproduces the `species' bound cutoff.


\section{$\AdS_5\times \IS^5$ orbifolds}
\label{sec:ads5}

In this section we consider type IIB string theory on $\AdS_5\times \IS^5/\IZ_k$. The discussion can be easily extended to general toric orbifold theories $\AdS_5\times \IX_5/\IZ_k$, but the 5-sphere case will suffice to illustrate the main points. We study general $\IZ_k$ actions compatible with supersymmetry, namely acting as $\SU(3)$ in the underlying $\IC^3$. We also note that, although these vacua do not display scale separation, we may discuss the $5$d physics essentially in the same sense as in the AdS/CFT correspondence, whose dictionary and results we use freely in this section. Moreover, our final statement involves gauge couplings for $\U(1)$ symmetries, which can be observed at arbitrarily long distances, in particular at energies well below the KK scale.

As pioneered in \cite{Gukov:1998kn} (see also \cite{Burrington:2006uu,Burrington:2006aw,Burrington:2006pu,Burrington:2007mj} for other examples) and generalized in \cite{Garcia-Valdecasas:2019cqn}, there is a discrete gauge symmetry in the $\AdS_5$ theory, corresponding to the discrete Heisenberg group ${\bf H}_k$. This is defined by  two non-commuting $\IZ_k$ symmetries generated by $A$, $B$ (hence $A^k=1$, $B^k=1$) satisfying
\beqa
AB=CBA\coma
\eeqa
with $C$ a central element (also generating a further $\IZ_k$, and possibly mixing with other anomaly free baryonic $\U(1)$'s, if present).

Generalizing \cite{Gukov:1998kn}, the particles charged under the discrete symmetry are D3-branes wrapped on torsion 3-cycles carrying non-trivial flat gauge bundles (discrete Wilson lines and 't Hooft loops). The minimally charged particle is obtained by wrapping the D3-brane on a maximal $\IS^3/\IZ_k$. We are interested in the mass of this particle, and in particular in its scaling with $k$. It is a simple exercise, as this is just analogous to a giant graviton in the parent $\AdS_5\times\IS^5$ theory \cite{McGreevy:2000cw}.

\bigskip

\paragraph{The D3-brane particle mass computation}\mbox{}

In the KK reduction from $10$d to $5$d, the $5$d Planck mass $M_{p,\,5}$ in terms of the string scale is
\beqa
M_{p,\,5}^{\,3}\, =\, \frac{M_s^8 R^5}{g_s^2\, k}\fstop
\label{5dplanck}
\eeqa
We are ignoring numerical factors e.g. in the volume of $\IS^5$. 
Above, $R$ is the curvature radius of $\IS^5$, which is also the $\AdS_5$ 
radius. Note that in order to get a theory with $N$ units of RR 5-form flux 
over $\IS^5/\IZ_k$, the parent theory is the $\AdS_5\times \IS^5$ solution 
corresponding to $Nk$ D3-branes, and the usual relation between the radius 
$R$ and $N$ is modified to
\beqa
R^4=4\pi (\alpha')^2 \, g_s\, N\, k\fstop
\eeqa
Hence
\beqa
R\sim M_s^{\,-1} g_s^{\,\frac 14 } \, N^{\frac 14}\, k^{\frac 14}\coma
\label{s5radius}
\eeqa
where we have dropped numerical factors.

The mass $m$ of the D3-brane particle\footnote{Notice that for our purposes it does not matter if we are in the string or Einstein frame, since this introduces factors that depend on dynamical fields, but does not change the scaling with $k$, which goes into the constant part (reference value).} in $5$d is
\beqa
m=\frac{M_s^4R^3}{g_s\, k}\fstop
\eeqa

We wish to express the mass in terms of the $5$d Planck scale. From (\ref{5dplanck}) and (\ref{s5radius}) we get
\beqa
M_s\sim M_{p,\,5} \, g_s^{\,\frac 14}\, N^{-\frac{5}{12}}\,k^{-\frac{1}{12}}\quad, \quad R\sim M_{p,\,5}^{\, -1}\, N^{\frac 23}\, k^{\frac 13}\fstop
\label{ms-r-scalings}
\eeqa
Hence
\beqa
m \sim M_{p,\,5} \, N^{\frac 13}\,k^{-\frac 13}\fstop
\label{mass-d3-planck}
\eeqa

Note that the $k$-dependence reproduces the $5$d version of the relation (\ref{dvali}) \cite{Dvali:2007hz}
\beqa 
m^3\, \sim \, \frac{M_{p,\,5}^{\, 3}}k\fstop
\label{dvali-5d}
\eeqa
This result fits nicely with the expectation for the mass of a particle charged under $\IZ_k$.

Notice that, as mentioned in Section \ref{sec:bh}, the coefficient in (\ref{dvali-5d}) is not necessarily the order of the discrete symmetry (which we recall is the Heisenberg group ${\bf H}_k$) but the number of particles emitted to peel off the black hole charge. We also note that the factor of $N$ in (\ref{mass-d3-planck}) is presumably related to the precise nature of the cutoff $\Lambda$ in the black hole argument in Section \ref{sec:dvali-wgc}. It would be interesting to explore this dependence in more detail, but we leave this for future work.

\bigskip

\paragraph{Comparison with the BPS formula and WCC}\mbox{}

The above states are not the lightest carrying charges under the  $\IZ_k$ subgroups of the Heisenberg group. In fact, there are charged particle states arising from fundamental strings and D1-branes wrapped on torsion 1-cycles on the internal geometry.
What is special about the above D3-brane particle states is that they are BPS. Just like giant gravitons in $\AdS_5\times \IS^5$, they carry $N$ units of momentum along a maximal $\IS^1$, determined by the $\IZ_k$ action. In the $5$d theory, there is a KK $\U(1)_R$, which is precisely the gravity dual of the R-symmetry of the holographic SCFT.  In the SCFT, the D3-brane particle states are dibaryons of the form $\det \Phi_{ij}$, with $\Phi$ denoting a generic bifundamental chiral multiplet in the quiver gauge theory. It has R-charge $N$, and conformal dimension $\Delta=N$. Using the AdS/CFT dictionary, we then expect the masses of these particles to be given by
\beqa
m=\frac{N}{R}\fstop
\eeqa
The fact that these states are BPS means that they should saturate the WGC conjecture bound, in other words, the BPS mass formula
\beqa
m\,=\,(g\,M_{p,\,5}^{\;\frac 12})\, NM_{p,\,5}\fstop
\label{mass-d3-wgc}
\eeqa
This is the standard $m=gQ$ in Planck units, with charge $q=N$ and $g$ being the gauge coupling of the $\U(1)$.

In these relations, there is no manifest dependence on $k$, which could be puzzling from the viewpoint of the black hole arguments. As we however know, the resolution is that, on these general grounds, the gauge coupling $g$ {\bf must} scale with $k$, at large $k$, in particular 
\beqa
g\sim k^{-\frac 13}\fstop
\eeqa

This is easily checked by computing the gauge coupling. In the KK reduction from $10$d to $5$d, the prefactor of the gauge kinetic term is
\beqa
\frac{1}{g^2}\,=\, \frac{M_s^8\,R^5}{g_s^2\, k}\, R^2\fstop
\eeqa
The first factor is just the $10$d prefactor times the volume of $\IS^5/\IZ_k$, and the $R^2$ comes from the rescaling of mixed components of the metric into dimensionful gauge field, such that charges are quantized in integers.

Using our above expressions, we get
\beqa
g\sim R^{-1}\, M_{p,\,5}^{\, -\frac 32}\coma
\eeqa
which means
\beqa
g\, M_{p,\,5}^{\;\frac12}\, =\, N^{-\frac 23}\, k^{-\frac 13}\fstop
\label{thegads5}
\eeqa
So, in terms of this gauge coupling, the mass (\ref{mass-d3-planck}) turns into (\ref{mass-d3-wgc}). Hence we recover a very explicit confirmation of our heuristic argument in Section \ref{sec:wcc}.

Let us conclude with some general remarks.
\begin{itemize}
	\item In addition to $\U(1)_R$ there are in general (in fact, for general toric theories) two extra mesonic $\U(1)$ symmetries, arising from isometries of the internal $5$d manifold. The direct computation of their $5$d gauge couplings proceeds as above, thus leading to a scaling compatible with the WCC.
	\item  In addition to D3-brane charged particles, there are  $5$d membranes of real codimension 2, which implement monodromies associated to the discrete group elements. As in the abelian case, these objects are charged under a dual discrete gauge symmetry (this can be made more manifest by introducing non-harmonic forms to represent the torsion classes \cite{Camara:2011jg,BerasaluceGonzalez:2012vb}). However, since these objects are not charged under any continuous symmetry, we lack a good handle to constrain their properties, and we will not discuss them further. 
\end{itemize}

\smallskip

\paragraph{The $\IZ_k$ Distance Conjectures}\mbox{}

It is interesting to explore the relation between the $\IZ_k$ WCC and the AdS Distance Conjecture in the present setup where, using (\ref{ms-r-scalings}), going to large $k$ implies going to large $R$. This is a decompactification limit (note that the orbifold only reduces lengths in $\IS^5$ in some directions, so the KK scale remains $R^{-1}$), in which also the AdS cosmological constant goes to zero, approaching flat space. Hence we can apply the AdS Distance Conjecture, which e.g. in its strong version (as we have supersymmetry) establishes that there should be a tower of states with masses scaling as
\beqa
m_{\rm tw} \sim \frac 1R \sim M_{p,\,5}\, N^{-\frac 23}\,k^{-\frac 13}\coma
\eeqa
where we have also kept the dependence on $N$. From the $1/R$ dependence, it is clear the tower corresponds to KK modes. These are the familiar particles dual to single trace chiral primary mesonic operators of the dual SCFT, extensively studied in the literature \cite{Witten:1998qj}, see \cite{Aharony:1999ti}. Note that, even though the scaling with $k$ is the same as for wrapped D3-branes, KK modes are lighter due to the relative factor of $N$.

\bigskip

\paragraph{A further subtlety}\mbox{}

The above discussion has overlooked an important subtlety. The discrete symmetry $\IZ_k$ (in fact the full discrete Heisenberg group) is intertwined with the $\U(1)$ in the following sense. Since the D3-branes are charged under the $\U(1)$ with charge $N$, a set of $k$ D3-branes carries no discrete $\IZ_k$ charge, but carries $kN$ units of momentum and cannot decay to the vacuum. In fact, the instanton processes removing the discrete $\IZ_k$ charge (which correspond to a D3-brane wrapped on the 4-chain whose boundary is $k$ times the torsion 3-cycle) produce simultaneously $N$ particles each carrying momentum $k$ on the circle (whose radius is $R/k$ due to the orbifold). 

The situation is very analogous to the one we will encounter in M-theory and type 
IIA compactifications in Section \ref{sec:ads4}, so we postpone the discussion. 
Suffice it to say that in this kind of situation, the actual discrete symmetry 
has order $k^2+N^2$, heuristically corresponding to the fact that the discrete 
charge may be eliminated via emission of $k$ D3-branes (each with charge $k$ 
under the discrete group) and $N$ KK modes (each with charge $N$ under the 
discrete group). In the regime where the gravity description of $\IS^5/\IZ_k$ 
is valid, we need large $R^4\sim Nk$  and large $R/k\sim N^{1/4}k^{-3/4}$, 
hence $N\gg k^3$, and the order of the gauge group is effectively dominated
by the $N^2$ term, corresponding to emission of $N$ KK modes. Hence, the actual 
discrete symmetry in this regime is an effective $\IZ_N$.

It is straightforward to repeat the above computations for the KK mode particles. The mass is given by $k/R$, as corresponds to mesonic operators of dimension $k$ (or multiples of it) due to the orbifold action. We obtain the relations and scalings
\beqa
m \,\sim \, M_{p,\,5}\, N^{-\frac 23} \, k^{\frac 23} \quad ,\quad m\sim g\, k\, M_{p,\,5}^{3/2} \quad ,\quad g\, M_{p,\,5}^{\;\frac12}\, =\, N^{-\frac 23}\, k^{-\frac 13}\fstop
\eeqa
Here $g$ is obviously the same as in \eqref{thegads5}, but we repeat it for convenience. Happily, it is clear that $g$ obeys a $\IZ_N$ WCC.
Notice also that the discretely charged KK modes fit more nicely with the black hole argument in Section \ref{sec:bh}. It seems more manageable to emit KK particles than D3-brane particles, as the later extend to a very large size in the internal dimension.

As anticipated, we will re-encounter a very similar situation in M-theory compactifications in the next section, with the additional handle of a type IIA reduction which makes these aspects far more intuitive. We refer the reader to those sections for details.

\section{M-theory orbifolds and ABJM}
\label{sec:ads4}

In this section we study the WCC in M-theory on $\AdS_4\times \IS^7/\IZ_k$ and its type IIA reduction, which provide the gravity dual of the ABJM gauge theories \cite{Aharony:2008ug}. These theories display interesting new subtleties as compared with earlier cases. Some have been partially discussed in the ABJM literature, so we can again profit from the holographic dictionary.

\subsection{M-theory on $\AdS_4\times \IS^7/\IZ_k$}
\label{sec:mtheory}

Let us now consider M-theory on $\AdS_4\times \IS^7/\IZ_k$, where $\IZ_k$ is generated by $z_i\to e^{2\pi i/k} z_i$ in the underlying $\IC^4$. This theory is the dual to the ABJM theories, which correspond to $\U(N)_k\times \U(N)_{-k}$ Chern-Simons matter theories,\footnote{Actually, as mentioned below and pointed out in \cite{Aharony:2008ug} the global structure is different such that there are gauge invariant dibaryons for arbitrary $N$, $k$.\label{foot:abjm}} with $\pm k$ denoting the CS level.

The curvature radius of the covering $\IS^7$ and the $\AdS_4$ are given by
\beqa
R^6\,=\, 2^5\pi^2 M_{p,\,11}^{\;-6} Nk\coma
\label{eq:RABJMMth}
\eeqa
where the factor of $Nk$ is analogous to that in Section \ref{sec:ads5}.

We are interested in studying gauge symmetries in the $4$d theory.  The $4$d Planck scale is given by
\beqa
M_{p,\,4}^{\;2}\, =\,\frac{M_{p,\,11}^{\;9} R^7}k\fstop
\eeqa
Hence we have
\beqa
R\, \sim\, M_{p,\,11}^{-1}\, N^{\frac 16} \, k^{\frac 16}\coma
\eeqa
and then
\beqa
M_{p,\,11}\, \sim\, M_{p,\,4}\, N^{-\frac 7{12}} \, k^{-\frac 1{12}}\quad ,\quad R\sim M_{p,\,4}^{\;-1} \, N^{\frac 34} k^{\frac 14}\fstop
\label{mp-r-scalings}
\eeqa

There are two relevant symmetries. There is a $\U(1)$ isometry, surviving from the underlying isometry of $\IS^7$ which decomposes as $\SO(8)\to \SU(4) \times \U(1)$ under the orbifold action $z_i\to e^{2\pi i/k} z_i$. It is a continuous gauge symmetry in $\AdS_4$. In addition, the internal space has a non-trivial torsion group $H_5\left(\IS^7/\IZ_k\right)=\IZ_k$ which allows to obtain $4$d particles by wrapping M5-branes on the torsion 5-cycle. In the covering space the minimal charge particle is essentially an M5-brane giant graviton, similar to those in the $\AdS_4\times \IS^7$ theory. In particular, it carries $N$ units of momentum on the $\IS^1$ associated to the $\U(1)$ symmetry.

This seems a perfect candidate for a WGC particle charged under the discrete symmetry, so we consider its properties, in analogy with the D3-brane particles in Section \ref{sec:ads5}. Its mass is given by
\beqa
m_{\rm M5}\sim \frac{M_{p,\,11}^{\;6} \, R^5}{k}\, =\, M_{p,\,4}\, N^{\frac 14} \, k^{-\frac 14}\coma
\eeqa
where, in the last equation, we have used (\ref{mp-r-scalings}). Note that we recover the AdS/CFT dictionary relation
\beqa
m_{\rm M5}\, = \, \frac NR\coma
\eeqa
indicating that the M5-brane particle is dual to an operator of conformal dimension $N$, as befits a dibaryon.

We can compare this mass with the WGC bound (BPS bound), by computing the gauge coupling. This is just given by the KK reduction of the $11$d Einstein terms and gives
\beqa
g^{-2}\,\sim \, M_{p,\,11}^{\;9} \, (\, R^7\, k^{-1} \,)\, R^2 \fstop
 \eeqa
Note that we have taken the normalization factor $R^2$, which holds when $\gcd(N,k)=1$. This is because in that normalization, the charges under the $\U(1)$ are KK modes of momentum multiple of $k$ (since the radius is $R/k$ due to the orbifold action), and M5-branes, whose charges are multiples of $N$. Then by Bezout's lemma, the minimal charge quantum is 1. For the general case $\gcd(N,k)=r$, we would have a factor $(R/r)^2$. We proceed with the coprime case in what follows. As pointed out in \cite{Aharony:2008ug}, the existence of gauge invariant dibaryon operators for general $N$ (not a multiple of $k$) implies a specific choice of the global structure of the gauge group of the holographically dual ABJM field theory, see footnote \ref{foot:abjm}.

Using (\ref{mp-r-scalings}) we have
\beqa
g^{-2}\, \sim \, N^{\frac 32}\, k^{\frac 12}\quad \rightarrow \quad g\, \sim\,N^{-\frac 34} \, k^{-\frac 14}\fstop
\label{mth-gauge-coupling}
\eeqa
So we get the WGC/BPS relation
\beqa
m_{\rm M5}\, =\, M_{p,\,4}\, g\, N\fstop
\label{eq:M5WGCABJM}
\eeqa

It is interesting that in the large $k$ limit we recover a weak coupling scaling
result $g\sim k^{-1/4}$, but that this decrease is slower  than the critical
$g\sim k^{-1/2}$ required by the black hole evaporation argument. The resolution
of this point reveals two interesting related subtleties: the actual discrete
gauge symmetry of the theory is not just $\IZ_k$, and the wrapped M5-branes
are not the only states charged under the discrete symmetry. Indeed, as
mentioned in \cite{Aharony:2008ug}, a set of $k$ wrapped M5-brane particles
can unwrap, but they do not decay to the vacuum, but rather turn into $N$  KK
states with momentum along the $\U(1)$ circle (which, due to the $\IZ_k$
orbifold, is quantized in multiples of $k$). In other words, there are
instantons (given by M5-branes wrapped on the $\IC\IP ^3$ base of the Hopf
fibration of $\IS^7/\IZ_k$), which emit $k$ M5-branes and $N$ minimal momentum
KK modes. As will be more intuitively  explained in Section \ref{sec:typeiia},
there is a discrete symmetry of order $N^2+k^2$, under which a wrapped
M5-brane has charge $k$ and a minimal momentum KK mode has charge $N$. Thus KK
modes provide a possible alternative to allow for black hole decay, which in
fact is dominated by processes of emission of $N$ such KK modes. Hence, the
gauge coupling needs to obey a WCC with respect to $N$. Let us thus check this
point.

The KK particle mass is given by
\beqa
m_{\rm KK}\, =\, \frac kR\fstop
\eeqa
This in fact constitutes the holographic dictionary relation for an operator of conformal dimension $k$. These are constructed with $k$ copies of a bifundamental field, as required by gauge invariance under the level-$k$ $\U(1)$'s of the holographic dual field theory \cite{Aharony:2008ug}.

Using (\ref{mp-r-scalings}) we have
\beqa
m_{\rm KK}\,=\, M_{p,\,4} \, N^{-\frac 34}\, k^{\frac 34}\coma
\eeqa
and with (\ref{mth-gauge-coupling}) we obtain 
\beqa
m_{\rm KK}\,=\, M_{p,\,4} \,g\, k\fstop
\label{eq:KKWGCABJM} 
\eeqa
Hence these are WGC particles charged under the discrete symmetry, and the gauge coupling \eqref{mth-gauge-coupling} obeys a WCC bound with respect to $N$.

\subsection{Type IIA description of ABJM vacua}
\label{sec:typeiia}

We may now describe the type IIA version of the previous section, which makes some of the above points more intuitive, and also provides a good warm-up for coming sections. 

The type IIA limit arises as follows. The $\IS^7$ is a $\IS^1$ Hopf fibration
over $\IC\IP ^3$, where the $\IZ_k$ quotient acts on the $\IS^1$. The radius of
the $\IC\IP ^3$ factor is large whenever $Nk\gg 1$. From~\eqref{eq:RABJMMth} we
conclude that the M-theory description is valid whenever $k^5\ll N$. When $k$
increases, we end up in a weakly coupled regime and we can reduce to type IIA
string theory \cite{Aharony:2008ug}.

The type IIA background corresponds to a compactification on AdS$_4\times \IC\IP ^3$ with internal and AdS radii $R_s$ (see below), with $N$ units of $F_6$ RR flux over $\IC\IP ^3$ (i.e. of $F_4$ flux over AdS$_4$) and $k$ units of $F_2$ RR flux over $\IC\IP ^1\subset \IC\IP ^3$ (due to the  Hopf fibration of the M-theory $\IS^1$).

The matching of string theory quantities to the $11$d Planck scale is as follows. The $10$d string coupling $g_s$ is related to the M-theory radius $\mathcal{R}=R/k$ as
\begin{equation}
g_s=M_{p,\,11}^{3/2}\,\mathcal{R}^{3/2}\coma
\end{equation}
that scales as
\begin{equation}
g_s\sim N^{\frac 14}\,k^{-\frac 54}\fstop
\label{abjm-dilaton}
\end{equation}
The string scale $M_s$ is related to the $11$d Planck scale as
\beqa
M_{p,\,11}^{\;3}\, =\frac{M_s^{\,3}}{g_s} \fstop 
\eeqa
So in terms of $M_s$, the radius \eqref{eq:RABJMMth} becomes
\begin{equation}
R
\sim N^{\frac 16}\,k^{\frac 16}\,g_s^{\frac 13}\,M_s^{-1}\fstop
\end{equation}

Finally we need the radius $R_s$ of $\IC\IP ^3$ from the string viewpoint. The type IIA metric is given by
\begin{equation}
ds_{IIA}^2=R_s^2\left(\frac{1}{4}ds^2_{\AdS_4}+ds^2_{\IC\IP ^3}\right)\coma
\end{equation}
where 
\begin{equation}
R_s^2
\,\sim\,
N^{1/2}\,k^{-1/2}\,M_s^{-2}\fstop
\label{eq:RsABJMIIA}
\end{equation}

We can now compute the $4$d Planck mass:
\begin{equation}
M_{p,\,4}^4\sim M_s^8\, g_s^{-2}\, R_s^6\coma
\label{eq:Mp4ABJMIIA}
\end{equation}
and combine with \eqref{eq:RsABJMIIA}, \eqref{eq:Mp4ABJMIIA} and \eqref{abjm-dilaton} to obtain
\begin{equation}
R_s \,\sim \, N\,k^{-1}\,M_{p,\,4}^{-1}\,g_s^{-1} \sim M_{p,\,4}^{\,-1} \,N^{\frac 34}\, k^{\frac 14}\quad \coma \quad M_s \,\sim\, N^{-3/4}\,k^{3/4} \, M_{p,\,4}\, g_s\,\sim\, M_{p,\,4}\, N^{-\frac 12} \,k^{-\frac 12}\fstop
\label{eq:RsMsMp4}
\end{equation}

Let us now consider the gauge symmetries in the $4$d theory in this type IIA
string compactification. The $\SU(4)$ symmetry arises as the isometry of the
internal $\IC\IP ^3$. On the other hand, there are additional $\U(1)$ gauge
fields arising from the $10$d RR fields, concretely the $10$d RR 1-form potential
and the $10$d RR 3-form potential integrated over $\IC\IP ^1\subset \IC\IP ^3$.
We should however notice that there are St\"uckelberg couplings arising from the
$10$d Chern-Simons coupling $B_2 F_2 F_6$, of the form\footnote{For further discussion of Chern-Simons couplings and swampland constraints see \cite{Montero:2017yja}.}
\beqa
N\, B_2\, F_2 \, +\, k\, B_2 F_2'\coma
\eeqa
where $F_2'=\int_{\IC\IP ^2} F_6$. This implies that the massless $\U(1)$ linear combination is 
\begin{equation}
J=kQ_0-NQ_4\fstop
\label{massless-J}
\end{equation} 
Here the generators $Q_0$, $Q_4$ are labeled by the objects charged under the corresponding $\U(1)$'s, namely D0-branes and D4-branes wrapped on $\IC\IP ^2$. Note that our sign convention differs from \cite{Aharony:2008ug}.

The orthogonal linear combination,
\beqa
Q_{\rm broken} \,=\, NQ_0+kQ_4\coma
\eeqa
corresponds to a massive $\U(1)$, which is broken by instanton effects, and only a discrete subgroup remains. The instanton corresponds to an NS5-brane wrapped on $\IC\IP ^3$, since it couples magnetically to $B_2$. It suffers from Freed-Witten anomalies due to the $F_6$ and $F_2$ fluxes, so it emits $N$ D0-branes and $k$ wrapped D4-branes. Hence, the total violation of $Q_{\rm broken}$ is $N^2+k^2$. This is the order of the gauge group. However, notice that at the level of the black hole (and of the WCC), what is actually relevant is the number of particles required to be emitted, namely $N$ D0-branes (contributing charge $N$ each) and $k$ D4-branes (contributing charge $k$ each). The type IIA internal space is large compared with the string scale if $N\gg k$, so the limit of large order of the discrete gauge group scales as $N^2$ and the black hole decay is dominated by the emission of $N$ D0-branes. In the arguments below, this is one particular instance in which the relevant coefficient in scaling relations is not the order of the discrete symmetry, but the number of emitted particles.

Notice also that we are recovering in possibly more intuitive terms the discussion of the earlier M-theory setup, with wrapped D4-branes corresponding to wrapped M5-branes and D0-branes corresponding to KK modes of the M-theory circle.

Let us discuss the masses of the D4- and D0-brane particles and the $\U(1)$ gauge couplings. They scale as
\beqa
&&m_{\rm D0}\, =\, g_s^{-1} \, M_s\, \sim M_{p,\,4}\, N^{-\frac 34} \,k^{\frac 
34}, \nonumber \\
&& m_{\rm D4}\, =\,g_s^{-1}\, M_s^{\, 5}\, R_s^{\,4}\, \sim M_{p,\,4}\, N^{\frac 14} \,k^{-\frac 14} \fstop
\eeqa
We already notice that the D0-brane mass decreases with $N$ faster than the `species' bound reviewed in Section \ref{sec:dvali-wgc}, ensuring that black holes can get rid of their discrete charge by emitting D0-branes. Let us turn to check the implication for gauge couplings and verify the $\IZ_N$ WCC.

The $4$d gauge couplings for the $\U(1)$'s generated by $Q_0$ and $Q_4$ are given by
\beqa
\frac{1}{g_0^{\,2}} &\sim &M_s^8\, R_s^6\, M_s^{-2} \nonumber \\
\frac{1}{g_4^{\,2}} &\sim &M_s^8\, R_s^6\, (\,M_s^{-5}\, R_s^{-4}\,)^{-2}\fstop
\eeqa
The first common factor arises from the reduction of the $10$d kinetic term for RR fields on the $\IC\IP ^3$, while the last factors arise from the normalization of the gauge fields by the coefficient of the D-brane Chern-Simons term, so that charges are integer numbers. Using the familiar relations above, we obtain the scalings
\begin{align}
g_{0}^{-2}\sim N^{3/2}k^{-3/2}\quad ,\quad
g_{4}^{-2}\sim N^{-1/2}k^{1/2}\fstop
\end{align}
The coupling constant associated to the massless combination \eqref{massless-J} is 
\begin{equation}
g^{-2}=\frac{k^2}{g_{0}^2}+\frac{N^2}{g_{4}^2}\sim N^{\frac 32}k^{\frac 12}\coma
\end{equation}
and, as explained, its scaling satisfies the WCC with respect to $N$
\beqa
g\sim N^{-\frac 34}k^{-\frac 14}\fstop
\eeqa

As expected, the D0- and D4-brane particles satisfy the BPS/WGC bound, in agreement with the result for wrapped M5-branes and KK modes in \eqref{eq:M5WGCABJM}, \eqref{eq:KKWGCABJM}
\beqa
m_{\rm D4}\, =\, M_{p,\,4} g\, N\quad ,\quad
m_{\rm D0}\,=\, M_{p,\,4} \,g\, k\fstop
\eeqa
Notice also that $g\sim 1/R$ in Planck units, so the above masses imply conformal dimensions $N$ and $k$ for the holographically dual operators, as is by now familiar.

\section{Discrete 3-form symmetries and scale separation in AdS solutions}
\label{sec:iiacy}

In \cite{Lust:2019zwm} it is proposed that in AdS vacua with cosmological constant $\Lambda$, the limit $\Lambda\to 0$  is accompanied by a tower of states becoming light as
\beqa
m\sim |\Lambda|^\alpha\fstop
\label{adc}
\eeqa
The strong version of this conjecture is that $\alpha=1/2$, which is the case in many/most string solutions (see below for examples). We focus on this version and phrase the conjecture as a ratio of scales\footnote{Note that $\Lambda$ has dimension mass${}^2$.}
\beqa
\frac{m^2}{\Lambda}\sim {\cal O}(1)\fstop
\label{strong-adc}
\eeqa
The states in the tower are typically KK states, and we use this term in the following. The conjecture implies that one cannot achieve a (parametric) separation of the KK scale and the scale of the cosmological constant. In fact, a problem that has been pervasive in holography literature is the search of gravity duals of QCD or $4$d SCFT with conformal anomaly coefficients $a\neq c$. Scale separation is also an important intermediate step in constructions attempting to realize de Sitter vacua in string theory \cite{Kachru:2003aw,Balasubramanian:2005zx}. Hence it is an important question which merits attention.

There are systematic constructions of AdS$_4$ vacua in string theory in type IIA compactifications on CY orientifolds with NSNS and RR fluxes \cite{DeWolfe:2005uu,Camara:2005dc} (see \cite{Marchesano:2019hfb} for a recent generalization to general CYs). As already noticed in the literature, there is a family of vacua in  \cite{DeWolfe:2005uu} (see also \cite{Camara:2005dc}) claimed to achieve scale separation, thus violating the strong form of the conjecture. In this section we show that this family enjoys a $\IZ_k$ discrete symmetry arising from 3-form gauge symmetries broken by a topological coupling to an axion, of the kind considered in \cite{Dvali:2005an,Kaloper:2008fb}, together with a continuous 3-form symmetry. Hence it provides a setup in which a $\IZ_k$ WCC for 3-form gauge fields is at work. The tension of the corresponding BPS domain walls can be related to the vacuum energy, and introduces additional factors of $k$ in (\ref{strong-adc}), thus explaining the parametric scale separation, that is controlled by the parameter $k$. This symmetry is consistently absent in other AdS vacua with no scale separation, hence provides a rationale for the existence of scale separation in this family, and suggests the proper generalization of (\ref{strong-adc}) in the presence of domain wall $\IZ_k$ symmetries.

\subsection{Review of scaling $\AdS_4$ vacua with scale separation}
\label{sec:dgkt}

In this section we review some key elements of the family of models with scale separation, following \cite{DeWolfe:2005uu} (see also \cite{Camara:2005dc} for related classes of type IIA AdS vacua).

Consider type IIA on a CY threefold modded out by an orientifold action introducing O6-planes. The O6-planes introduce a tadpole for the RR 7-form, which is canceled by (possibly present) D6-branes, and a combination of the $F_0\equiv m$ Romans mass flux parameter and $H_3$ NSNS field strength flux on 3-cycles. Although it is possible to introduce it, we consider the RR $F_2$ field strength fluxes to be zero.\footnote{Actually, by monodromies in suitable axions \cite{Marchesano:2014mla} the $F_2$ flux can be generated due to the presence of $F_0$ flux. This follows from a Dvali-Kaloper-Sorbo coupling, and intertwines non-trivially with similar DKS coupling to appear in Section \ref{sec:dks}. We keep our simplified discussion for $F_2=0$, and refer the reader to \cite{Herraez:2018vae,Escobar:2018tiu} for further information on the more general framework.\label{monosubtle}} On the other hand, we introduce RR $F_4$ field strength fluxes on a basis of 4-cycles ${\tilde \Sigma}_i$
\beqa
\int_{\tilde \Sigma_i} \, F_4\, =\, e_{\tilde i}\, \in \IZ\fstop
\eeqa
We do not introduce RR $F_6$ flux over the CY, and only consider it when generated by monodromies, see Section \ref{sec:scaling-from-symms}.
Some details on the $4$d effective action of this theory are provided in Appendix \ref{app:gaugeDGKT}, and here we streamline the key facts. 
Whereas the fluxes $F_0=m$ and $H_3$ are constrained to be ${\cal O}(1)$ due to the tadpole conditions, the fluxes for $F_4$ are unconstrained and can be taken large. The scaling solutions are achieved in the large $k$ limit of
\beqa
e_{\tilde i} \, \sim \, {\bar e}_{\tilde i}\, k\coma
\eeqa
where the ${\bar e}_{\tilde i}$ are ${\cal O}(1)$ quantities. Note that we have renamed the scaling parameter of \cite{DeWolfe:2005uu} as $k$ to make better contact with earlier sections, and to emphasize its forthcoming role as related to a discrete gauge symmetry.

Although we keep much of the upcoming discussion general, it is useful to consider explicit examples. A simple class is obtained by taking toroidal orbifolds $\IT^6/\IZ_3$, whose untwisted sector is given by 3 K\"ahler moduli associated to the 3 underlying $\IT^2$'s. Their volumes, measured in string units, are denoted by  $v_i$, $i=1,2,3$, with the overall volume being ${\bar {\cal {V}}}\sim v_1v_2v_3$.  They are complexified by the axions from the NSNS 2-form over the 2-tori $b_i$. We ignore twisted sectors, and refer the reader to \cite{DeWolfe:2005uu} for details. Since $h_{2,1}=0$, there is only one axion $\xi$ from the period of the RR 3-form over the 3-cycle; it combines with the $4$d dilaton $e^D$ to form a complex modulus.

In the scaling limit, \cite{DeWolfe:2005uu} found a supersymmetric AdS$_4$ minimum (which we refer to as the DGKT solution) with the following values for the $4$d moduli
\beqa
v_i,b_i\, \sim\, k^{\frac 12} \quad ,\quad {\bar {\cal {V}}}\sim k^{\frac 32}\quad  ,\quad e^{-D}, \xi \, \sim \,k^{\frac 32}\fstop
\label{fundamental-scalings}
\eeqa
This implies that
\beqa
M_s^{\, 2} \sim e^{2D} M_{p,\,4}^{\,2}\sim k^{-3} M_{p,\,4}^{\,2}\coma
\label{ms-scalings}
\eeqa
and that the following relevant quantities of the $4$d effective action, evaluated at the minimum, and measured in $4$d Planck units, scale as
\beqa
W\sim k^{\frac 32}\quad ,\quad e^{\cal{K}}\sim k^{-\frac{15}{2}}\quad \Lambda\sim k^{-\frac 92}\fstop
\label{effective-scalings}
\eeqa

One may evaluate the KK scale  as
\beqa
m_{\rm KK}\,\sim \, {\bar{\cal V}}^{-\frac 16} \, M_s\sim k^{-\frac 74} \, M_{p,\,4}
\label{mkk-scalings}
\eeqa
(incidentally, it coincides with the mass scale for other massive moduli, so it provides a general cutoff of the $4$d theory).

This leads to a relation of the type \eqref{adc} 
\beqa
m_{\rm KK}^{\; 2}\sim \Lambda^{\frac 79}\coma
\eeqa
and hence to a seeming parametric violation of the strong version of the conjecture.
In \cite{Font:2019uva} the problem was considered in a family of IIA compactifications with geometric fluxes. The back-reaction of the latter \cite{Aldazabal:2007sn} implied a modification of $m_{\rm KK}$ which restored the scaling predicted by the strong AdS Distance Conjecture. This mechanism however is not obviously available in the present context, where geometric moduli are absent. In the following sections we propose the scale separation is physical in these cases, and find a rationale in terms of underlying symmetries.

\subsection{The discrete 3-form symmetry}
\label{sec:dks}

In this section we address the backbone of the solution to the above conundrum. First, notice that we had rewritten the strong conjecture as in the form (\ref{strong-adc}) with hindsight. Indeed, taking this ratio we find that in the DGKT family
\beqa
\frac{m_{\rm KK}^{\, 2}}\Lambda\sim k\fstop
\eeqa
Alternatively, we may express the vacuum energy $\Lambda$ in terms of the UV cutoff scale $m_{\rm KK}$ as
\beqa
\Lambda\sim \frac{m_{\rm KK}^{\, 2}}k\fstop
\label{scale-separation-guay}
\eeqa
Recalling that $\Lambda$ has dimension 2, this is extremely reminiscent of the type of relation one finds in theories with a $\IZ_k$ discrete gauge symmetry, see \eqref{dvali}. Moreover, since the left hand side quantity is the vacuum energy, the relevant charged objects should be related to the structure of the vacuum. 

We now show that there is indeed an effective $\IZ_k$ symmetry acting on domain walls changing the fluxes in the vacuum. The structure is controlled by topological couplings of the $10$d theory. In fact, we will study them without assuming the vacuum solution described in the previous section, and show that the scaling relations found there are a consequence of these topological couplings, or equivalently of the discrete symmetry structure.

So we start with the general CY (orientifold) compactification, and consider the basis of 4-cycles ${\tilde \Sigma_i}$ and their dual 2-cycles $\Sigma_i$. We recall the $F_4$ flux structure and introduce $4$d axions from $B_2$ as
\beqa
\int_{\tilde \Sigma_i} F_4\,=\, k\,{\bar e}_{\tilde i}\quad ,\quad \int_{ \Sigma_i} B_2\,=\,\phi_i
\eeqa
(these axions were denoted by $b_i$ in the toroidal setup above). In addition, we introduce a symplectic basis of orientifold-odd 3-cycles $\alpha_a$ and orientifold-even 3-cycles $\beta_a$, and introduce the NSNS $H_3$ fluxes and RR axions 
\beqa
\int_{\alpha_a} H_3\,=\, p_a\quad ,\quad\int_{\beta_a} C_3=\xi_a\fstop
\label{h3flux-dgkt}
\eeqa
In addition, there is a Romans mass flux parameter $F_0=m$.

Let us initially focus on the dynamics of K\"ahler moduli, hence ignore $\xi_a$, which will be reintroduced later on. Most of the discussion is general, although we eventually apply it to the toroidal orbifold for illustration.

The dimensional reduction of the $10$d Chern-Simons coupling $F_4F_4B_2$ leads to the $4$d topological coupling
\beqa
k\, \left(\, \sum_i\, {\bar e}_{\tilde i} \phi_i\, \right) F_4\fstop
\eeqa
This makes the 3-form massive, by eating up the 2-form dual to a linear combination of axions. The overall factor $k$ implies that there is a discrete $\IZ_k$ symmetry under which domain walls are charged \cite{BerasaluceGonzalez:2012zn}. This confirms we are on the right track. In fact, although certain modifications are about to come in, in the large $k$ limit this $\IZ_k$ discrete symmetry determines the properties of the system.

The situation is actually slightly more subtle, because of the following. The scalars $\phi_i$ also appear in couplings with other 4-forms, arising from the 8-form as
\beqa
F_{4,\,{\tilde i}}\,=\,\int_{\tilde \Sigma_i} F_8\fstop
\eeqa
Hence, including the reduction of the $10$d coupling $F_0B_2F_8$, the complete set of topological couplings is
\beqa
\, m\, \sum_i\, \phi_i F_{4,\,{\tilde i}}\,+ k\,\left (\, \sum_i \,{\bar 
e}_{\tilde i} \phi_i\right) F_4\fstop
\label{full-4form-ks}
\eeqa 
This means that the combination $\phi'\equiv \sum_i {\bar e}_{\tilde i}\phi_i$ also couples to other 4-forms. To isolate that dependence, introduce the generators $Q'$  and $Q_i$ of 3-form $\U(1)$ symmetries  for $C_3$ and $C_{3,i}$, and consider the linear combination
\beqa
Q'\,=\, \sum_i\, {\bar e}_{\tilde i} Q_i\fstop
\eeqa
The topological coupling for the corresponding 4-form $F_4'$ is
\beqa
m \,\left(\sum_i\, {\bar e}_{\tilde i}\, \phi_i \,\right)\, F_4'\, = \,m\,\phi'\, F_4'\fstop
\eeqa
Hence, we can isolate  the axion $\phi'$ with its couplings to the 4-forms $F_4$, $F_4'$ as
\beqa
\phi'\, \left( m\,F_4' \, +\, k\, F_4 \right)\fstop
\label{universal-4form-sector}
\eeqa
It is interesting that we have this universal sector, decoupled (at the topological level) from other axions and 4-forms, and hence independent of the details of the underlying CY compactification space.

Since there is only one axion and two 4-forms, there is clearly a massless 3-form corresponding to the combination
\beqa
Q_{\U(1)} \, =\, k\, Q'\, -\, m\,Q\, =\, \sum_i e_{\tilde i} \,Q_{\tilde i} \,-\,m\,Q\fstop
\label{dw-cont-sym}
\eeqa
In the second equality we have recast the combination in terms of the original 4-forms. It is straightforward to check, using (\ref{full-4form-ks}), that $Q_{\U(1)}$ is indeed free from topological couplings to scalars, hence remains an unbroken 3-form gauge symmetry.

The combination appearing in (\ref{universal-4form-sector}), namely
\beqa
Q_{\perp}\, =\, mQ'+kQ\, =\sum_i \, m\, {\bar e}_{\tilde i} \, \, Q_{\tilde i}\,+\,k\,Q\coma 
\eeqa
is broken to a discrete subgroup. To better understand its structure, consider the string emitting a number of domain walls, and let us compute the violation of conservation of $Q_{\perp}$. The relevant string couples to the dual to $\phi$, namely it is given by an NS5-brane wrapped on the linear combination of  4-cycles $\sum_i {\bar e}_{\tilde i}{\tilde \Sigma}_i$. Due to the presence of $m$, it emits $m{\bar e}_{\tilde i}$ D6-branes wrapped on ${\tilde\Sigma}_i$; due to the presence of $e_{\tilde i}$ units of 4-form flux over ${\tilde \Sigma}_i$, it emits $\sum_i {\bar e}_{\tilde i} e_{\tilde i}$ D2-branes. Since each D6-brane on  ${\tilde\Sigma}_i$ violates $Q_{\tilde{i}}$ in 1 unit, and each D2-brane violates $Q$ in 1 unit, we have a total violation of $Q_{\rm broken}$ by
\beqa
\Delta Q_{ \perp} \,=\, \sum_i \left({\bar e}_{\tilde i}\right)^2 \, \left(k^2+m^2\right)\fstop
\eeqa
Although it would seem that at large $k$ the symmetry is of order $k^2$, notice that it suffices to have $k$ D2-branes (plus a number of D6's sub-leading in the $1/k$ approximation) to annihilate into a string. It's only that one D2-brane implies a violation of $k$ units of $Q_{\rm broken}$, from the way we built the linear combination. So it is an effective $\IZ_k$ for D2-branes.

Notice that this system realizes a 3-form version of the theories with discrete and continuous $\U(1)$ symmetries (for 1-forms) we described in earlier sections. In particular, the structure of two underlying $\U(1)$'s with one linear combination broken by a topological coupling is completely analogous to the discussion of the type IIA gravity dual of ABJM theories in Section \ref{sec:typeiia}.\footnote{With the notational difference that the roles of $N$, $k$ are now played by $k$, $m$, respectively.}

\subsection{Scaling relations for moduli from discrete symmetries}
\label{sec:scaling-from-symms}

In analogy with the ABJM system, the D2- and D6-brane domain walls are BPS, and their tensions must relate to their charges under the unbroken $Q_{\U(1)}$, 
\beqa
T_{\rm DW}\, =\, g\, Q_{\U(1)} \, M_{p,\,4}^{\,4}\fstop
\label{bps-dw}
\eeqa
The gauge coupling $g$ for $Q_{\U(1)}$ is derived from those of the 3-form symmetries associated to $Q$ and $Q_{\tilde i}$, see \eqref{dw-cont-sym}. We denote them $g_2$, $g_{6,\,{\tilde i}}$  respectively, to indicate that the charged objects are D2-branes and D6-branes on $\Sigma_{\tilde i}$. We have
\beqa
\frac{1}{g^2}\, =\,  k^2\, \sum_i\,\left({\bar e}_{\tilde i}\right)^2\, \frac{1}{g_{6,\,\tilde i}^{\,2}}\, +\, m^2\, \frac{1}{g_2^2}\fstop
\label{gauge-coupling-linearcomb}
\eeqa
The fact that both D2- and D6-branes can satisfy the BPS condition (\ref{bps-dw}), implies that, in the large $k$ limit, their gauge couplings must relate as
\beqa
g_{6,\,{\tilde i}}\,\sim\, k\,g_2\fstop
\label{scaling-ratio-gauge-couplings}
\eeqa
It is easy to express the ratio of these gauge couplings in terms of microscopic compactification parameters and {\em derive} that the scaling for $v$ reproduces \eqref{fundamental-scalings}. We offer a simplified discussion here, referring the reader to Appendix \ref{app:gaugeDGKT} for a supergravity-friendly derivation. For concreteness, we also focus on the toroidal case. The inverse gauge couplings squared are
\beqa
&&\frac{1}{g_2^{\;2}}\, =\, M_s^{\, 2} {\bar{\cal V}} \, \left(\, M_s^{-3}\,\right)^2 \, =\, M_s^{\, -4}\, \bar{\cal V}\coma\nonumber \\
&&\frac{1}{g_{6,\,{\tilde i}}^{\,2}}\, =\, M_s^{\, 2} {\bar{\cal V}} \, \left(\, M_s^{-3}\, \frac{v_i}{\bar{\cal{V}}}\,\right)^2\, =\, M_s^{\, -4}\,\frac{(v_i)^2}{\bar{\cal{V}}}\coma
\label{gauge-couplings-26}
\eeqa
where the first factor arises from the $10$d coupling and the terms in parenthesis arise from normalization of charges to integers, and we recall that ${\bar{\cal V}}=v_1v_2v_3$. We have that
\beqa
\frac{g_{6,\,{\tilde i}}}{g_2}\, =\, \frac{\bar{\cal V}}{v_i}
\eeqa
and comparing with \eqref{scaling-ratio-gauge-couplings} for different $i$'s gives
\beqa
v_i\,\sim\, k^{\frac 12}\quad ,\quad {\bar{\cal{V}}}\,\sim\, k^{\frac 32}\fstop
\label{scaling-vis}
\eeqa

A more direct, and possibly more general, route to the scaling relations for moduli is to use the monodromy relations. The fact that e.g. $F_4$ has topological couplings to axions implies that the flux $N$ of $F_6$ over the CY changes as the axions wind across their periods.
Indeed, the above discussion is slightly oversimplified, since the fluxes experience a more intricate set of axion monodromies. These have been studied systematically in \cite{Herraez:2018vae}, and appeared implicitly in \cite{DeWolfe:2005uu}. They just follow from the nested structure of $10$d Chern-Simons terms, or equivalently of the $10$d modified Bianchi identity for $F_6$, which implies
\beqa
F_6 \,=\, dC_5\, +\, F_4 B_2\, +\, F_2\, B_2\, B_2\, +\, F_0\,  B_2\, B_2\, B_2\, + H_3 C_3\fstop
\eeqa
Hence, restricting to our setup with only $F_0$, $F_4$ and $H_3$, the effective $4$d theory can depend only on the combination
\beqa
N\, +\, k\,{\bar e}_{\tilde i} \, \phi_i\, +\, m\, \kappa_{ijk} \, \phi_i\phi_j\phi_k\, +\, p_a\, \xi_a
\label{monodromy-dgkt}
\eeqa
(where sums over repeated indices are implicit). Here $\kappa_{ijk}$ is the triple intersection number. For instance, $\kappa_{123}=1$ for the torus. This implies that it is possible to generate $F_6$ flux from $m$ by performing a monodromy in $b_1$ to generate $F_2$ on the first $\IT^2$, followed by a monodromy in $b_2$ to generate $F_4$ on the $\IT^4$ transverse to the third coordinate, and one in $b_3$ to generate $F_6$ on the CY. 

This is a more complete version of the topological  couplings to 4-forms we have been considering, and which underlies the discrete symmetry of the system. We are interested in its behavior in the large $k$ limit. Consistent scaling of the monodromy relations for large $k$ requires that
\beqa
\phi_i\, \sim\, k^{\frac 12}\fstop
\eeqa
This is the generalization of the scaling for $b_i$ in \eqref{fundamental-scalings}, and provides the complexified counterpart of our scalings for $v_i$ in \eqref{scaling-vis} (which recovered those in \eqref{fundamental-scalings}). We point out that the fact that the two components of complex moduli have identical scalings with large flux quanta fits nicely with results on asymptotic flux compactification \cite{Grimm:2019ixq}. It is extremely interesting that this result follow from just the discrete symmetry in the present context.

Motivated by this, we can use a similar argument to extract the scaling of the dilaton multiplet in the large $k$ limit. From \eqref{monodromy-dgkt} we get
\beqa
\xi_a\,  \sim \, k^{\frac 32}\fstop
\eeqa
This is the complexification of a similar dependence of the dilaton, which thus reproduces \eqref{fundamental-scalings}. 

Interestingly, with this information, which in particular implies the scaling \eqref{ms-scalings}, i.e. $M_s\sim k^{-3/2} M_{p,\,4}$, we obtain the scaling of gauge couplings \eqref{gauge-couplings-26}, \eqref{gauge-coupling-linearcomb}
\beqa
g_2\, \sim\, k^{-\frac{15}4} \quad ,\quad g_{6,\,{\tilde i}}\, \sim \, k^{-\frac{11}{4}}\quad , \quad g\,\sim\, k^{-\frac{15}4}\coma
\eeqa
providing a nice version of the WCC for domain walls.
 
Note however that when including the $H_3$ fluxes, the above discussion is equivalent to the inclusion of additional topological couplings $p_a\xi_aF_4$. In other words, D2-brane domain walls, in the presence of $H_3$ flux, can annihilate in sets of $p_a$ by nucleating a string given by a D4-brane wrapped on the 3-cycle $\alpha_a$, due to the Freed-Witten inconsistency of the latter. The presence of these couplings spoils the structure of continuous and discrete 3-form gauge symmetries  found in the K\"ahler moduli sector. In other words, the coupling of $F_4$ to a different linear combination of axions implies that the former continuous symmetry is actually also broken by the new additional axion, given by the linear combination of $\xi_a$. We skip the detailed discussion of the resulting complete discrete symmetry group. Note however that for large $k$ the effects of both $m$ and $p$ are sub-leading in a $1/k$ expansion, so the $\IZ_k$ symmetry we have been using prevails.

Since we have recovered the scalings of the K\"ahler and complex structure moduli, it is a simple exercise to use the expressions of $4$d supergravity to derive others like \eqref{effective-scalings}, and eventually recover the scale separation \eqref{scale-separation-guay}. On the other hand, the $4$d approach has been criticized as potentially hiding subtleties of the $10$d solution. Therefore in the following we use an alternative approach, and exploit properties of BPS domain walls to recover the vacuum energy.


\subsection{Discrete symmetries and scale separation}
\label{sec:domain-walls}

In this section we exploit the interplay between the tensions of domain walls and the vacuum energy, and study the interplay of discrete symmetries and scale separation. We argue through explicit examples that AdS vacua with trivial discrete symmetry for domain walls  do not have scale separation; this is true even if there are non-trivial discrete symmetries for particles or strings, and in general for real codimension higher than 1 objects. On the other hand, we show that the above type IIA modes with non-trivial discrete symmetry for domain walls, with the corresponding scaling for moduli, do have vacuum energy with scale separation. We extend this general relation and put forward the following refined version of the  swampland constraint \eqref{strong-adc}, as follows:

\bigskip

\paragraph{$\IZ_k$ Refined Strong AdS Distance Conjecture:} {\em  Consider 
quantum gravity on an AdS vacuum with a $\IZ_k$ discrete symmetry for domain 
walls (with $k$ large). In the flat-space limit $\Lambda\to 0$ (with $\Lambda 
k\to 0$ as well) there exists an infinite tower of states at a scale $M_{\rm 
cutoff}$, with the relation}
\beqa
\Lambda \, \sim \, \frac{M_{\rm cutoff}^2}{k}\fstop
\eeqa

We now proceed to check this conjecture in the examples of supersymmetric AdS vacua of this paper, by deriving their vacuum energies from the properties of domain walls.
%


\subsubsection{Vacuum energy from domain walls}

Let us describe our main tool to evaluate the vacuum energies without invoking an underlying scalar potential. There is in fact a general relation between domain wall tensions and vacuum energies, which essentially follows from junction conditions in general relativity. We refer the reader to Appendix \ref{app:RSjunction} for a discussion well adapted to our application in AdS. The key point is that the domain wall tension $T$ is the variation of certain quantities $\lambda$, see \eqref{thevs}, whose square essentially gives the vacuum energy $\Lambda$, see \eqref{thelambdas}. In the supersymmetric setup, and for BPS domain walls, these statements become the familiar
\beqa
\lambda\, =\, e^{{\cal K}/2} \, W\quad , \quad  T\,=\, \Delta(e^{{\cal K}/2}\,W) =\, \Delta\lambda\,\quad ,\quad \Lambda\, =\, -3e^{\cal K}\, |W|^2\sim -|\lambda|^2\fstop
\eeqa
 
We consider BPS domain walls whose quantized charge describes the change in some field strength flux $n$ as one crosses the domain wall. In the limit of large flux $n$, the tension $T$ provides the derivative of $d\lambda/dn$. We can then solve to obtain the scaling with $n$ of $\lambda$, and thus of its square, $\Lambda$.

\subsubsection{Warm-up examples: no scale separation}

We now turn to discuss the AdS examples of Sections \ref{sec:ads5}, \ref{sec:ads4}, {\em deriving} their AdS radius from the above strategy, and showing there is no scale separation. This is in agreement with our Refined Strong AdS Distance Conjecture (RSADC), as these examples have discrete symmetries for particles (and for their dual real codimension 2 objects) but not for domain walls. 

\bigskip

\paragraph{Type IIB on $\IS^5/\IZ_k$}\mbox{}

Consider type IIB on $\IS^5/\IZ_k$ with $N$ units of RR 5-form flux and 
\beqa
R^4 \sim M_s^{-4}\, g_s\, N\,k\fstop
\eeqa
This is of course the class of theories considered in Section \ref{sec:ads5}, but we are now not imposing the solution for the $5$d vacuum, rather we are deriving its vacuum energy from the domain wall properties. In passing, we also discuss the gauge coupling of the 3-forms and draw conclusions regarding the WCC.

We consider a BPS domain wall given by  a D3-brane in $5$d. Its tension is
\beqa
T_{\rm D3} \, \sim \, M_s^{\,4}\, g_s^{-1}\, \sim\, M_{p,\,5}^{\,4} N^{-\frac 53} k^{-\frac 13}\fstop
\eeqa
The same result is obtained from the BPS condition
\beqa
T_{\rm D3}=g Q_{\rm D3}
\eeqa
upon computation of the gauge coupling of the $5$d RR 4-form under which the D3-brane is charged. Since the tension essentially agrees with the gauge coupling, we observe an interesting WCC scaling for $g$ (in that respect, recall that the relevant large order discrete symmetry is $\IZ_N$). This is interesting, since the discrete symmetry acts on particles/membranes, whereas $g$ is a 3-form gauge coupling. It would be interesting to explore the interplay between discrete and continuous symmetries of different degrees; we hope to come back to this in future work.

Since this domain wall interpolates among vacua with $N$ and $N+1$, one can now obtain
\beqa
\frac{d\lambda}{dN}\,\sim N^{-\frac 53} k^{-\frac 13} \quad \Rightarrow\quad \lambda\, \sim \, N^{-\frac 23}\, k^{-\frac 13}
\quad \Rightarrow\quad
\Lambda \sim M_{p,\,5}^{\,2}\,N^{-\frac 43}\, k^{-\frac 23}\fstop
\eeqa
Using (\ref{ms-r-scalings}) we have
\beqa
\Lambda\sim R^{-2}\fstop
\eeqa
Hence the AdS radius is the same as that of the internal space, and there is no decoupling of scales. This is the strong ADC statement in~\cite{Lust:2019zwm}. 

Note that, even though there are discrete gauge symmetries in the system, their orders do not enter the ratio of scales. This is in agreement with our RSADC, since these discrete symmetries involve particles and membranes, not domain walls.

\bigskip

\paragraph{M-theory on $\IS^7/\IZ_k$}\mbox{}

Let us consider M-theory on $\IS^7/\IZ_k$ with $N$ units of flux (or $Nk$ in the covering space) and
\beqa
R^6\, \sim\, M_{p,\,11}^{-6} \, N\, k\fstop
\eeqa
This is of course the same system as in Section \ref{sec:ads4}, but again we wish to {\em derive} the $4$d vacuum energy from the relevant BPS domain walls.  We consider a BPS domain wall given by an M2-brane in $4$d. Its tension is
\beqa
T_{\rm M2}\, \sim \, M_{p,\,11}^3\,\sim\,M_{p,\,4}^{\;3}\, N^{-\frac 74}\,k^{-\frac 14}\fstop
\eeqa
where we used (\ref{mp-r-scalings}). The same result is obtained from the BPS condition
\beqa
T_{\rm M2}\, =\, g\, Q_{\rm M2}
\eeqa
upon computation of the gauge coupling $g$ for the $4$d 3-form. Recalling the relevant large order discrete symmetry is $\IZ_N$, we note again that we get an interesting WCC scaling for $g$.

Since the M2-brane domain wall interpolates between vacua with $N$ and $N+1$ units of flux, we have
\beqa
\frac{d\lambda}{dN}\,\sim\, N^{-\frac 74}\,k^{-\frac 14}\quad \Rightarrow\quad 
\lambda \, \sim\, N^{-\frac 34}\, k^{-\frac 14} \quad \Rightarrow\quad \Lambda\,\sim\, M_{p,\,4}^{\,2}\, N^{-\frac 32} \,k^{-\frac 12}\,\sim\, R^{-2}\fstop
\eeqa
In the last relation, we have used (\ref{mp-r-scalings}).  Again, we recover the result that the AdS radius is of the same order of magnitude as the KK scale of the internal space. Also, notice that there are discrete symmetries in the theory, but they involve particles and strings, rather than domain walls. Hence, they do not alter the relation between scales, in agreement with our RSADC.

\bigskip

 \paragraph{Type IIA on $\IC\IP ^3$}\mbox{}
 
We would like to repeat the previous computation in the type IIA picture. Let us consider type IIA theory on $\IC\IP ^3$ with $N$ units of $F_6$ RR flux over $\IC\IP ^3$ and $k$ units of $F_2$ RR flux over $\IC\IP ^1\subset \IC\IP ^3$ and 
\begin{equation}
R_s^2 \sim M_s^{-2} N^{1/2}k^{-1/2}\fstop
\end{equation}
This is the same system as in Section~\ref{sec:typeiia}. The relevant BPS domain wall is  a D2-brane in $4$d, whose tension is 
\begin{equation}
T_{\rm D2}\,\sim\, M_s^3\, g_s^{-1}\, \sim\, M_{p,4}^3 N^{-7/4}k^{-1/4}\fstop
\end{equation}
This is the same scaling as the M2-brane in the previous section, and the D2-brane domain wall interpolates vacua with $N$ and $N+1$ units of flux, so we recover
\begin{equation}
\Lambda \sim M_{p,4}^2 N^{-3/2}k^{-1/2} \sim R_s^{-2}\fstop
\end{equation}
The $\AdS$ radius is the same as that of the internal space, with no scale separation, in agreement with our RSADC.

\medskip

\subsubsection{Revisiting the Scale Separation in type IIA CY flux compactifications}

Consider now the configurations with the large $k$ discrete $\IZ_k$ symmetry for domain walls in Section \ref{sec:dks}. We wish to derive the scaling of the vacuum energy with $k$, just using the scaling of moduli vevs \eqref{fundamental-scalings}, \eqref{ms-scalings} derived in Section \ref{sec:scaling-from-symms} from the $\IZ_k$ symmetry. 

We consider the BPS domains wall given by a  D4-brane wrapped on the combination of 2-cycles ${\sum_i \bar e}_{\tilde i}\Sigma_i$. This domain wall interpolates between vacua with $F_4$ flux given by $k$ and $k+1$. Notice that the $F_4$-flux is not monodromic, hence the D4-branes are stable against nucleation of strings, and can provide BPS objects (in contrast with e.g. D2- and D6-brane domain walls encountered in earlier sections).

The tension of these domain walls can be obtained from the BPS equation and the gauge couplings, computed in detail in Appendix \ref{app:gaugeDGKT}. Here we carry out a simplified derivation, taking the toroidal case for concreteness. The gauge coupling of a D4$_i$-brane domain wall is
\beqa
\frac{1}{g_{4,\,i}^{\;2}}\, =\, M_s^2\, {\bar{\cal V}} \, (\, M_s^{\,-3}\,v_i^{-1}\, )^2\, =\, M_s^{-4}\, {\bar{\cal V}} v_i^{-2}\, \sim \, k^{\frac{13}{2}}\fstop
\eeqa
As usual, in the first equality, the first term comes from the reduction of the $10$d coupling, and the parenthesis from the charge normalization. Note that the scaling is common for all $i$, so by the BPS condition we get the tension
\beqa
T_{\rm DW}\sim k^{-\frac{13}4}\fstop
\eeqa
Notice that, if interpreted in terms of gauge couplings, this implies an interesting WCC, as in earlier examples. From the above tension we get
\beqa
\frac{d\lambda}{dk}\,\sim\, k^{-\frac{13}4}\quad \Rightarrow \quad 
\lambda\,\sim\, k^{-\frac 94}  \quad \Rightarrow \quad \Lambda\, \sim 
\,k^{-\frac 92}\fstop
\eeqa
So we recover the scaling \eqref{effective-scalings} for $\Lambda$ (the reader 
can check those of ${\cal K}$ and $W$ as well). Once $m_{\rm KK}$ is recovered 
as in \eqref{mkk-scalings}, this reproduces the scale separation 
\eqref{scale-separation-guay}, in agreement with our RSADC conjecture.

\hspace*{5cm}
%
\section*{Acknowledgments}
We are pleased to thank L. Ib\'a\~nez, F. Marchesano  for useful discussions. This work is supported by the Spanish Research Agency (Agencia Espa\~nola de Investigaci\'on) through the grant IFT Centro de Excelencia Severo Ochoa  SEV-2016-0597, and the grant
 FPA2015-65480-P from the MCU/AEI/FEDER. The work by J.C. is supported by a FPU position from Spanish Ministry of Education.  A.M. received funding from ``la Caixa" Foundation (ID 100010434) with fellowship code LCF/BQ/IN18/11660045 and from the European Union’s Horizon 2020 research and innovation programme under the Marie Sk\l odowska-Curie grant agreement No. 713673.
\newpage

\appendix

\section{Species bound for extremal black holes}
\label{app:extremal}

In the following we consider the evaporation of extremal black holes endowed
with $\IZ_k$ charge. For concreteness, the classical solutions we are taking are
the extremal Reissner-Nordstr\"{o}m black holes in $4$d space-time dimensions.
They have vanishing Hawking temperature, so the analysis in \cite{Dvali:2007hz}
is not directly applicable.

Extremal black holes can discharge through Schwinger radiation
\cite{PhysRev.82.664,PhysRevD.41.1142,gibbons_vacuum_1975}. Whenever the
electric field is much larger than the background curvature, this happens
essentially in flat space \cite{Montero:2019ekk}. In this case the production
rate has an exponential suppression
\begin{equation}\label{production-rate}
    \Gamma \sim e^{-\frac{m^{2}}{qE}} \sim e^{-\chi}\,,
\end{equation}
where $ m $ and $ q $ are the mass and charge of the emitted particle and $ E $ 
is the electric field, given by
\begin{equation} \label{electric}
E = \frac{g^2}{4\pi} \frac{Q}{r^{2}}\,.
\end{equation}

As argued in Section \ref{Zk-WGC}, the simplest way in which this kind of black
hole is able to get rid of both continuous and discrete charge while remaining
sub-extremal is in the presence of a $ \IZ_k $ WGC particle. Let us assume that
this particle is actually BPS,
\begin{equation}\label{BPS-4d}
    m = g q M_{p}\,.
\end{equation} 
As a consequence, the black hole will remain extremal throughout the whole 
evaporation process. 

From \eqref{production-rate} and \eqref{electric}, we notice that the maximum
particle production will happen close to the horizon, so in this order of
magnitude analysis we will approximate the whole radiation as the contribution
of that region.

From the extremality condition we can relate the horizon radius
and the charge of the BH with its mass through
\begin{equation}
    r_{h} \sim \frac{M_{BH}}{M_{p}^{2}}, \qquad gQ \sim \frac{M_{BH}}{M_{p}}\,.
\end{equation}
They lead to
\begin{equation}\label{electric-field}
    E \sim g \frac{M_{p}^{3}}{M_{BH}}\,.
\end{equation}

Introducing (\ref{BPS-4d}) and (\ref{electric-field}) in 
(\ref{production-rate}) we can estimate the factor in the exponential 
suppression of the production rate of the $ \IZ_k $ WGC particle to be
\begin{equation}\label{chi-final}
    \chi \sim \frac{m M_{BH}}{M_{p}^{2}}\,.
\end{equation}
The black hole will be able to efficiently evaporate discrete charge when
\begin{equation}\label{efficient-evap}
    M_{BH} \lesssim \frac{M_{p}^{2}}{m}\,.
\end{equation}

With this condition being true, the black hole should still have enough mass to
radiate $\mathcal{O} \left( k\right)$ particles (assuming the $ \IZ_k $ WGC
particle to have unit discrete charge), which means
\begin{equation}\label{Zk-evap}
    M_{BH} \gtrsim k m\,.
\end{equation}
Finally, from the two conditions (\ref{efficient-evap}) and (\ref{Zk-evap}), we obtain the following bound for the mass of the $ \IZ_k $ WGC particle:

\begin{equation}
    m^2 \lesssim \frac{M_{p}^{2}}{k}\,.
\end{equation}
This is the species bound in \cite{Dvali:2007hz}. We have shown that the bound 
also applies to extremal black holes emitting $ \IZ_k $ WGC particles via 
Schwinger effect.

\section{Discrete symmetries in intersecting brane models}
\label{app:intersecting}

Discrete symmetries are ubiquitous in models of intersecting branes (see  \cite{Ibanez:2012zz} for a review), as pioneered in \cite{BerasaluceGonzalez:2011wy}. In this appendix we use them to illustrate the interplay of $\IZ_k$ and $\U(1)$ gauge symmetries, and the scalings implied by the $\IZ_k$ WCC.

Let us start by recalling the basic setup. Consider a compactification of type IIA  on a Calabi-Yau space $\IX_6$ quotiented by the orientifold\footnote{Note that the orientifolds are not essential for the argument, but we choose to introduce them to better connect with the literature on intersecting brane models.} action $\Omega {\cal R}(-1)^{F_L}$, where ${\cal R}$ is an antiholomorphic $\IZ_2$ involution of $\IX_6$, which introduces O6-planes. Let us denote $[\Pi_{\rm O6}]$ the total homology class of the 3-cycles wrapped by the O6-planes. Introducing a symplectic basis $[\alpha_i]$, $[\beta_i]$ of 3-cycles even and odd under ${\cal R}$, respectively, we may expand 
\beqa
[\Pi_{\rm O6}]\, =\, \sum_i r_{\rm O6}^i [\alpha_i] \, +\, s_{\rm O6}^i [\beta_i]\coma
\eeqa
with $r_{\rm O6}^i$, $s_{\rm O6}^i$ some coefficients of order 1-10.

The O6-planes are charged under the RR 7-form, so to cancel its tadpoles we introduce D6-branes. We consider stacks of $N_A$ overlapping D6$_A-$branes wrapped on 3-cycles $\Pi_A$, and their orientifold image D6$_{A'}-$branes on 3-cycles $\Pi_{A'}$. In terms of the basis, we have
\beqa
[\Pi_A]\, =\,\sum_i r_{A}^i [\alpha_i] \, +\, s_{A}^i [\beta_i]\quad, \quad [\Pi_{A'}]\, =\,\sum_i r_{A}^i [\alpha_i] \, -\, s_{A}^i [\beta_i]\fstop
\eeqa

The RR tadpole condition reads
\beqa
\sum_A 2r_A^i + r_{\rm O6}^i\, =\, 0 \quad \forall i\fstop
\eeqa
In addition there are K-theory RR tadpole conditions \cite{Uranga:2000xp}, which we skip in this sketchy discussion.

In these models, there are St\"uckelberg couplings for the $\U(1)_A$, of the form
\beqa
\sum_A\, N_A s_A^i b_{2,i} \, F_A\coma
\eeqa
where wedge product is implicit. $F_A$ is the field strength of the $\U(1)$ 
gauge field on the D6$_A-$branes, and the $4$d 2-forms $b_{2,i}$ arise from the 
KK compactification of the RR 5-form $C_5$ as
\beqa
b_{2,i}\, =\, \int_{\beta_i} C_5\fstop
\eeqa
This makes some of the $\U(1)$'s massive. Let us consider linear combinations of the $\U(1)_A$ generators $Q_A$
\beqa
Q\, =\, \sum_A c_A Q_A\coma
\label{u1comb}
\eeqa
with $c_A$ being coprime integers, so as to preserve charge integrality.
The St\"uckelberg coupling for the field strength $F$ of the $\U(1)$ generated by $A$ is
\beqa
\left(\, \sum_A \, c_A N_A s_A^i\, \right)\, b_{2,i}\, F\fstop
\eeqa
Hence, the condition for a $\U(1)$ to remain massless is
\beqa
\sum_A \, c_A N_A s_A^i\,=0\quad \forall i\fstop
\label{masslessu1cond}
\eeqa
If not, the $\U(1)$ is broken, remaining only as approximate global symmetry, broken by non-perturbative D2-brane instanton effects \cite{Blumenhagen:2006xt,Ibanez:2007rs,Florea:2006si}. The condition that a discrete $\IZ_k$ subgroup remains as exact discrete gauge symmetry is
\beqa
\sum_A \, c_A N_A s_A^i\,=\,0\; {\rm mod}\; k \quad \forall i\fstop
\eeqa
Generically, to achieve this for large $k$ a possibility\footnote{This is not the only one, but we stick to it as an illustrative example.} is to have $s_A^i\sim k$, at least for some $A$, for all $i$. This implies that there is some brane which is wrapped on a very large (i.e. multiply wrapped) cycle. This implies that in general any unbroken $\U(1)$, given by a linear combination (\ref{u1comb}) satisfying (\ref{masslessu1cond}), will also involve that particular $Q_A$ with a coefficient of order $k$. This implies that the gauge coupling of the unbroken $\U(1)$ scales as
\beqa
\frac{1}{g^2} \,=\, k\quad {\rm hence} \; g\sim k^{-\frac 12}\coma
\eeqa
in agreement with the $\IZ_k$ WCC.

Although this is not quite a rigorous argument, it is a good illustration of how the interplay between $\U(1)$ gauge couplings and $\IZ_k$ symmetries arises, as a consequence of the fact that, to achieve a large order $\IZ_k$ discrete symmetry, one needs to use parametrically large cycles, thus parametrically scaling gauge couplings to zero. Hence, intersecting brane models provide an intuitive mechanism for the $\IZ_k$ WCC. More detailed string theory examples are presented in the main text.

\section{Gauge couplings in type IIA CY compactifications}
\label{app:gaugeDGKT}

In this appendix we derive the gauge coupling constants for domain walls present in type IIA CY flux compactifications. We review the computation in~\cite{Font:2019cxq} following the conventions in \cite{DeWolfe:2005uu}.
From~\cite{DeWolfe:2005uu}, the $10$d string frame action is given by\footnote{Our convention is that $|F_p|^2=F_{\alpha_1\ldots \alpha_p}F^{\alpha_1\ldots \alpha_p}/p!$.}
\begin{equation}
S^{10d}=\frac{1}{2\kappa_{10}^2}\int d^{10}x\sqrt{-g}\left(e^{-2\phi}(R+4(\de_\mu\phi)^2-\frac{1}{2}|H_3^{\text{total}}|^2)-(|\tilde{F}_2|^2+|\tilde{F}_4|^2+m_0^2)\right)+S_{CS}\coma
\end{equation}
where $2\kappa_{10}^2=(2\pi)^7\alpha'^4$ and the definitions of the field strengths are
\beqa
&&H_3^{\text{total}}=dB_s+H_3^{\text{bg}}\coma\nonumber \\
&&\tilde{F}_2=dC_1+m B_2\coma\\
&&\tilde{F}_4=dC_3+F_4^{\text{bg}}-C_1\wedge H_3-\frac{m}{2}B_2\wedge B_2\fstop\nonumber
\eeqa
The Chern-Simons action contains also a prefactor $(2\kappa_{10})^{-1}$ in front. We define an adimensional internal volume by  $\bar{\mathcal{V}}=M_s^6\mathcal{V}$ and perform the dimensional reduction in the string frame. For instance, the kinetic term for the $4$d field strength associated to the $10$d $F_4$ reads
\begin{equation}
S_{4d}^{\text{kin}} \supset- \frac{M_s^2}{2}\int d_4 x\sqrt{-g_4}\bar{\mathcal{V}}|\mathcal{F}_4|^2 \coma
\end{equation}
where $\mathcal{F}_4=dC_3$. To move back to the Einstein frame, we choose a reference scale $a$, and define the $4$d dilaton $D(x)$ as
\begin{equation}
a=\frac{\langle \bar{\mathcal{V}}\rangle}{e^{2\langle\phi\rangle}} \quad , \quad 
e^{2D}=\frac{e^{2\phi}}{\bar{\mathcal{V}}}\fstop
\end{equation}
So the Einstein frame kinetic terms take the form
\begin{equation}
S_{E}^{\text{kin}}\supset\frac{M_s^2}{2a}\int d^4x\sqrt{-g_E}R_E-\frac{a^2M_s^2}{2}\int d^4x\sqrt{-g_E}\bar{\mathcal{V}}e^{-4D}|\mathcal{F}_4|^2\coma
\label{eq:SEkinaction}
\end{equation}
where the products are now done using $g_E$ as a metric. 

To obtain $4$d gauge 3-forms, we perform a KK reduction of $10$d p-forms along suitable harmonic $(\text{p}-3)$-forms in the internal space. In the notation of~\cite{Herraez:2018vae}, 
\begin{equation}
C_3=c_3^0 \coma C_5=c_3^a \wedge \omega_a \coma C_7=\tilde{d}_{3a}\wedge \tilde{\omega}^a \text{ and } C_9=\tilde{d}_3\wedge \omega_6\fstop
\label{eq:Cp1decom}
\end{equation}
They corresponds to the relevant 4-forms $\mathcal{F}_4^0$, $\mathcal{F}_4^a$, 
$\tilde{\mathcal{F}}_{4,a}$ and $\tilde{\mathcal{F}}_4$ associated to D2-, D4-, 
D6- and D8-branes.\\

Notice that we need to normalize the gauge fields by the coefficient in front of the D-brane Chern-Simons term, in order for the charges to be properly quantized. For a Dp-brane this introduces factors of $\mu_p\propto\alpha'^{(p+1)/2} \sim M_s^{(p+1)}$ in the forthcoming gauge couplings. Namely, in order to be consistent, we need to keep the harmonic forms as adimensional, so the generic Chern-Simons action is
\begin{equation}
S^{(p)}_{CS}\sim M_s^{3}\int_{W_3\times \gamma_{p-2}} c_3 \wedge \omega_{p-2}\coma
\end{equation}
where we have called $c_3\wedge\omega_{p-2}$ collectively each decomposition in \eqref{eq:Cp1decom}. The normalization consists in redefining the RR 3-form by a factor $M_s^3$, so that there is no prefactor in front of the Chern-Simons action. The effect of such redefinition on~\eqref{eq:SEkinaction} is just a change in the prefactor in front of the kinetic terms of the gauge fields,
\begin{equation}
S_{E}^{\text{kin}}\supset\frac{M_s^2}{2a}\int 
d^4x\sqrt{-g_E}R_E-\frac{a^2}{2M_s^2}\int 
d^4x\sqrt{-g_E}\bar{\mathcal{V}}e^{-4D}|\mathcal{F}_4|^2\fstop
\label{eq:SEkinaction2}
\end{equation}
We are almost done in the definition of the coupling constants, but first we need the following quantities:
\beqa
M_s^2 \propto \frac{e^{2\langle\phi\rangle}}{\langle\bar{\mathcal{V}}\rangle}M_{p,\,4}^2=a^{-1}M_{p,\,4}^2\coma \quad
K_K =-\ln(8\bar{\mathcal{V}})\coma\quad
K_Q=4D\coma\quad
K=K_K+K_Q\fstop\nonumber
\eeqa
Substituting in \eqref{eq:SEkinaction2} and including the other 4-forms, we obtain~\cite{Font:2019cxq}
\begin{equation}
S^{kin}_E=\frac{\pi}{2M_p^4}\int \frac{e^{-K}}{8}\left[\mathcal{F}_4^0\wedge\star \mathcal{F}_4^0+4g_{ab}\mathcal{F}_4^a\wedge\star \mathcal{F}_4^b+\frac{1}{4\bar{\mathcal{V}}^2}g^{ab}\tilde{\mathcal{F}}_{4|a}\wedge\star \tilde{\mathcal{F}}_{4|b}+\frac{1}{\bar{\mathcal{V}}^2}\tilde{\mathcal{F}}_4\wedge\star \tilde{\mathcal{F}}_4\right]\coma
\end{equation}
where 
\begin{equation}
g_{ab}=\frac{\de^2 K_K}{\de t^a \de \bar{t}^b}
\end{equation}
is the metric in the K\"ahler moduli space with $t^a=v^a+ib^a$.\\
We need now to specialize to the toroidal orbifold in~\cite{DeWolfe:2005uu}. The K\"ahler potential is
\begin{equation}
K_K=-\ln(8v^1v^2v^3)=-\ln((t^1+\bar{t}^1)(t^2+\bar{t}^2)(t^3+\bar{t}^3))\coma
\end{equation}
so 
\begin{equation}
g_{ab}=\frac{1}{4}\diag\left((v^1)^{-2},(v^2)^{-2},(v^3)^{-2}\right)\fstop
\end{equation}
We rewrite the action according to this metric obtaining
\begin{equation}
S^{kin}_E=\frac{\pi}{2M_p^4}\int \frac{e^{-K}}{8}\left[\mathcal{F}_4^0\wedge\star \mathcal{F}_4^0+\sum_{i=1}^3\left(\frac{1}{(v^i)^2}\mathcal{F}_4^i\wedge\star \mathcal{F}_4^i+\frac{(v^i)^2}{\bar{\mathcal{V}}^2}\tilde{\mathcal{F}}_{4|i}\wedge\star \tilde{\mathcal{F}}_{4|i}\right)+\frac{1}{\bar{\mathcal{V}}^2}\tilde{\mathcal{F}}_4\wedge\star \tilde{\mathcal{F}}_4\right]\fstop
\end{equation}
We are finally able to read the coupling constants of all kinds of domain walls:
\beqa
&&\frac{1}{g_0^2}=\frac{\pi e^{-K}}{8M_p^4}\coma\quad 
\frac{1}{g_i^2}=\frac{\pi e^{-K}}{8M_p^4 (v^i)^2}\coma\nonumber \\
&&\frac{1}{g_{\tilde{i}}^2}=\frac{\pi e^{-K}  (v^i)^2}{8 M_p^4\bar{\mathcal{V}}^2}\coma\quad
\frac{1}{g_{\tilde{4}}^2}=\frac{\pi e^{-K}}{8M_p^4\bar{\mathcal{V}}^2}\fstop
\eeqa
From the main text, the scalings with the flux $k$ are
\begin{equation}
e^{K}\sim k^{-15/2} \coma v_i \sim k^{1/2} \e \bar{\mathcal{V}}\sim k^{3/2}\coma
\end{equation}
so the couplings scale as
\beqa
&&\frac{1}{g_0^2}=\frac{\pi e^{-K}}{8M_p^4}\sim k^{15/2}\quad ,\quad
\frac{1}{g_i^2}=\frac{\pi e^{-K}}{8M_p^4 (v^i)^2}\sim k^{13/2}\coma\nonumber \\
&&\frac{1}{g_{\tilde{i}}^2}=\frac{\pi e^{-K}  (v^i)^2}{8 M_p^4\bar{\mathcal{V}}^2}\sim k^{11/2}\quad,\quad
\frac{1}{g_{\tilde{4}}^2}=\frac{\pi e^{-K}}{8M_p^4\bar{\mathcal{V}}^2}\sim k^{9/2}\fstop
\eeqa

\section{Junction conditions for $\AdS$ vacua}
\label{app:RSjunction}

Here we adapt  to the $4$d setup the discussion of~\cite{Hatanaka:1999ac}, which studies a  Randall-Sundrum construction~\cite{Randall:1999ee,Randall:1999vf} with an arbitrary number of branes (domain walls). The discussion is also similar to systems of D8-branes in type I' theory \cite{Polchinski:1995df}.

Consider a $4$d spacetime with $N$ parallel domain walls with tensions $T_i$, located at positions $y_i$ in a coordinate $y$. The region between the $i^{th}$ and $(i+1)^{th}$ brane has cosmological constant $\Lambda_i$. A solution of the $4$d Einstein equations
\begin{align}
\sqrt{-G}
\left( R_{MN} - \frac{1}{2}G_{MN} R \right)
=&
-\frac{1}{4M_{p,\,4}^2}
\left[
\sum_{i=1}^N 
\Lambda_i \, [ \theta(y-y_i) - \theta(y-y_{i+1}) ]
\sqrt{-G}\,G_{MN}+
\nonumber \right.\\
& +
\left.\sum_{i=1}^N T_i \sqrt{-g^{(i)}} g^{(i)}_{\mu\nu} 
\delta^\mu_M \delta^\nu_N \delta(y-y_i)
\right] \label{eq:Einstenequ}
\end{align}
is given by the ansatz

\begin{equation}
ds^2 = e^{-2\sigma(y)}\eta_{\mu\nu} dx^\mu dx^\nu + r_c^2 dy^2\fstop
\label{eq:dssigma}
\end{equation}
The warp factor in the above expression is given by the following piecewise linear function 
\begin{align}
\sigma(y)
=& (\lambda_1 - \lambda_0) (y-y_1) \theta(y-y_1) 
+(\lambda_2 - \lambda_1)(y-y_2) \theta(y-y_2)+
\nonumber \\
&+ \ldots 
+(\lambda_N-\lambda_{N-1})(y-y_N)\theta(y-y_N)\coma
\label{eq:sigmasol}
\end{align}
where $\lambda_0$ and $\lambda_N$ provide the asymptotic behavior at $y\mp \infty$.
In any region between two domain walls, we can perform a change of coordinates
\begin{equation}
\frac{x_0}{r_c}=e^{\sigma(y)}\coma
\end{equation}
to bring the metric~\eqref{eq:dssigma} to a more standard form, i.e.

\begin{equation}
ds^2=\frac{r_c^2}{x_0^2}\left(\eta_{\mu\nu} dx^\mu dx^\nu+dx_0^2\right).
\end{equation}
from which it is clear that the solution describes slices of AdS$_4$ with different values of the cosmological constant, made explicit below.

From~\eqref{eq:Einstenequ}, we obtain the following constraints for $\sigma(y)$~\cite{Hatanaka:1999ac}:
\begin{align}
(\sigma'(y))^2&=-\frac{r_c^2}{12M_{p,\,4}^2}\sum_{i=1}^N\Lambda_i\left[\theta(y-y_i)-\theta(y-y_{i+1})\right]\coma\label{eq:sigmaprime}\\
\sigma''(y)&=\frac{r_c}{8M_{p,\,4}^2}\sum_{i=1}^NT_i\delta(y-y_i)\fstop\label{eq:sigmasecond}
\end{align}
Substituting \eqref{eq:sigmasol} in \eqref{eq:sigmaprime} and~\eqref{eq:sigmasecond}, we obtain the relations
\beqa
&& \lambda_i=\pm \sqrt{\frac{-\Lambda_ir_c^2}{12M_{p,\,4}^2}} \label{thelambdas}\coma \\
&&  \frac{T_ir_c}{8 M_{p,\,4}^2}=\lambda_i-\lambda_{i-1} \fstop \label{thevs}
\eeqa
Hence these junction conditions relate the variation of the cosmological constant to the potential of the branes that give us the domain walls. This is a general interpretation of what we proposed in Section~\ref{sec:domain-walls}.

\newpage

\bibliographystyle{JHEP}
\bibliography{mybib}

\end{document}